\catcode`\@=11
\newif\if@fewtab\@fewtabtrue

{\count255=\time\divide\count255 by 60
\xdef\hourmin{\number\count255}
\multiply\count255 by-60\advance\count255 by\time
\xdef\hourmin{\hourmin:\ifnum\count255<10 0\fi\the\count255}}
\def\ps@draft{\let\@mkboth\@gobbletwo
    \def\@oddhead{}
    \def\@oddfoot
       {\hbox to 7 cm{$\scriptstyle Draft\ version:\ \draftdate$
       \hfil}\hskip -7cm\hfil\rm\thepage \hfil}
    \def\@evenhead{}\let\@evenfoot\@oddfoot}

\def\ceqno{\global\@fewtabfalse
    \ifcase\@eqcnt \def\@tempa{& & &}\or \def\@tempa{& &}
      \or \def\@tempa{&}
      \or\def\@tempa{}\fi\@tempa
{\rm(\theequation)}}

\def\aeqno#1{\global\@fewtabfalse
    \ifcase\@eqcnt \def\@tempa{& & &}\or \def\@tempa{& &}
      \or \def\@tempa{&}
      \or\def\@tempa{}\fi\@tempa
{\rm(\theequation,#1)}}

\def\label#1{\ifnum\draftcontrol=1
 \global\def\draftnote{$\scriptstyle #1$}\fi
 \@bsphack\if@filesw {\let\thepage\relax
   \def\protect{\noexpand\noexpand\noexpand}%
\xdef\@gtempa{\write\@auxout{\string
      \newlabel{#1}{{\@currentlabel}{\thepage}}}}}\@gtempa
   \if@nobreak \ifvmode\nobreak\fi\fi\fi
  \@esphack}

\def\alabel#1#2{\label{#1}\global\@fewtabfalse
    \ifcase\@eqcnt \def\@tempa{& & &}\or \def\@tempa{& &}
      \or \def\@tempa{&}
      \or\def\@tempa{}\fi\@tempa
{\hbox to 3cm{\phantom{\rm(\theequation,#2)}
\draftnote \hfil}\hskip -3cm {\rm(\theequation,#2)}}}

\def\clabel#1{\label{#1}\global\@fewtabfalse
    \ifcase\@eqcnt \def\@tempa{& & &}\or \def\@tempa{& &}
      \or \def\@tempa{&}
      \or\def\@tempa{}\fi\@tempa
{\hbox to 3cm{\phantom{\rm(\theequation)}
\draftnote \hfil}\hskip -3cm{\rm(\theequation)}}}

\def\eqnarray{\def\draftnote{{}}\global\@fewtabtrue
\stepcounter{equation}\let\@currentlabel=\theequation
\global\@eqnswtrue
\global\@eqcnt\z@\tabskip\@centering\let\\=\@eqncr
$$\halign to \displaywidth\bgroup\@eqnsel\hskip\@centering\@eqcnt\z@
  $\displaystyle\tabskip\z@{##}$&\global\@eqcnt\@ne
  \hskip 1\arraycolsep \hfil${##}$\hfil
  &\global\@eqcnt\tw@ \hskip 1\arraycolsep
$\displaystyle\tabskip\z@{##}$
\hfil  \tabskip\@centering&\global\@eqcnt\thr@@\llap{##}\tabskip\z@
\cr}

\def\endeqnarray{\@@eqncr\egroup
      \global\advance\c@equation\m@ne$$\global\@ignoretrue}

\def\@eqnnum{\hbox to 3cm{\phantom{\rm(\theequation)} \draftnote
			 \hfil}\hskip -3cm {\rm(\theequation)}}

\def\@@eqncr{\let\@tempa\relax
    \ifcase\@eqcnt \def\@tempa{& & &}\or \def\@tempa{& &}
      \or \def\@tempa{&}
      \or\def\@tempa{}
\fi\@tempa
\if@eqnsw
\if@fewtab\@eqnnum\fi
\stepcounter{equation}\fi\global
\@eqnswtrue\global\@eqcnt\z@\global\@fewtabtrue\cr}

\def\draftcite#1{\ifnum\draftcontrol=1#1\else{}\fi}

\def\@lbibitem[#1]#2{\item{}\hskip -3cm \hbox to 2cm
{\hfil$\scriptstyle\draftcite{#2}$}\hskip
1cm[\@biblabel{#1}]\if@filesw
     {\def\protect##1{\string ##1\space}\immediate
      \write\@auxout{\string\bibcite{#2}{#1}}}\fi\ignorespaces}

\def\@bibitem#1{\item\hskip -3cm \hbox to 2cm
{\hfil $\scriptstyle\draftcite{#1}$}\hskip 1cm
\if@filesw \immediate\write\@auxout
       {\string\bibcite{#1}{\the\value{\@listctr}}}\fi\ignorespaces}

\def\nsection#1{\section{#1}}

\font\tendl=msbm10  scaled \magstep1
\font\sevendl=msbm7 scaled \magstep1
\font\fivedl=msbm5 scaled \magstep1
\font\tengl=eufm10  scaled \magstep1
\font\sevengl=eufm7 scaled \magstep1
\font\fivegl=eufm5 scaled \magstep1

\newfam\dlfam \def\dl{\fam\dlfam\tendl} 
\textfont\dlfam=\tendl \scriptfont\dlfam=\sevendl
\scriptscriptfont\dlfam=\fivedl
\newfam\glfam  
\textfont\glfam=\tengl \scriptfont\glfam=\sevengl
\scriptscriptfont\glfam=\fivegl

\def\draftdate{\number\month/\number\day/\number\year\ \ \ \hourmin }

\global\def\draftcontrol{0}
\catcode`\@=12
\def\tilde{\widetilde}
\def\hat{\widehat}
\def\pref#1{(\ref{#1})}
\documentstyle[11pt]{article}

\def\theequation{\arabic{equation}} 

\newcommand{\be}{\begin{eqnarray}}
\newcommand{\en}{\end{eqnarray}\vs 0.5 cm}
\newcommand{\non}{\nonumber}

\newcommand{\vs}{\vskip}
\newcommand{\hs}{\hspace}

\newcommand{\un}{\underline}
\newcommand{\NC}{{{\dl C}}}
\newcommand{\NZ}{{{\dl Z}}}
\newcommand{\NOm}{{{\bf\Omega}}}
\newcommand{\qq}{\begin{eqnarray}}
\newcommand{\de}{\bar\partial}
\newcommand{\da}{\partial}
\newcommand{\ee}{{\rm e}}

\newcommand{\qqq}{\end{eqnarray}}

\newcommand{\la}{\lambda}

\newcommand{\ga}{\gamma}

\newcommand{\tr}{\hbox{tr}}

\newcommand{\CA}{{\cal A}}

\newcommand{\CG}{{\cal G}}

\newcommand{\CM}{{\cal M}}
\newcommand{\CN}{{\cal N}}
\newcommand{\CO}{{\cal O}}
\newcommand{\CP}{{\cal P}}

\newcommand{\CT}{{\cal T}}
\newcommand{\CU}{{\cal U}}

\newcommand{\CW}{{\cal W}}

\newcommand{\Nh}{{\bf h}}
\newcommand{\Ng}{{\bf g}}
\newcommand{\Nn}{{\bf n}}
\newcommand{\s}{\hspace{0.05cm}}

\newcommand{\lle}{\langle}
\newcommand{\rle}{\rangle}
\begin{document}
\title{Unitarity of the Knizhnik-Zamolodchikov-Bernard connection
and the Bethe Ansatz for the elliptic Hitchin systems}
\author{\ \\Fernando Falceto \\ Depto. F\'{\i}sica
Te\'orica, Univ. Zaragoza, E-50009 Zaragoza, Spain
\\ \\Krzysztof Gaw\c{e}dzki \\ I.H.E.S., C.N.R.S.,
F-91440  Bures-sur-Yvette, France}
\date{ }
\maketitle

\vskip 0.3cm

\begin{abstract}
We work out finite-dimensional integral formulae for the
scalar product of genus one states of the group $G$ Chern-Simons
theory with insertions of Wilson lines.
Assuming convergence of the integrals,
we show that unitarity of the elliptic
Knizhnik-Zamolodchikov-Bernard connection with respect
to the scalar product of CS states is closely related
to the Bethe Ansatz for the commuting Hamiltonians
building up the connection and quantizing the quadratic
Hamiltonians of the elliptic Hitchin system.
\end{abstract}
\vskip 2cm

\nsection{Introduction}
\vskip 0.5cm

The present paper continues the program
\cite{Quadr}\cite{FGK}\cite{G}\cite{G1}
aimed at analysis of the scalar
product of states in the Chern-Simons (CS) theory.
It extends the considerations of ref.\s\s\cite{FG} were
we treated the $SU(2)$ CS theory on the elliptic curve
(times the time-line and with insertions of time-like
Wilson lines) to the case
of a general group $G$. As in the previous papers
of the series, the point is to express
the formal scalar product of the CS theory, given
by a functional integral over gauge fields, as
a multiple finite-dimensional integral. The latter,
if convergent for every state, provides
the space of CS states $W$ with a Hilbert space structure
and the holomorphic vector bundle $\CW$, obtained by varying
the modulus $\tau$ of the elliptic curve and
the positions $z_n$ of insertions,
with a hermitian structure.
\vskip 0.2cm

The integral expressions for the scalar product
of the CS states are close cousins of the contour
integral expressions for the conformal blocks
of the corresponding WZW conformal theory.
In the elliptic case, the contour integral
representations were recently studied
in ref.\s\s\cite{FV}. Our approach elucidates
the origin of the complicated expressions
which appear in such representations
for a general group: they are induced by
a simple trick, already used in \cite{FGK},
which handles a change of variables
in the functional integral.
\vskip 0.2cm

The WZW conformal
blocks are holomorphic sections $\theta$ of the bundle
$\CW$ of the CS state spaces satisfying in the elliptic
case the Knizhnik-Zamolodchikov-Bernard (KZB) equations
\cite{KZ,B1}
\qq
(\s\da_\tau\s+\s{_1\over^\kappa}\s H_0\s)\s\theta\s=\s0\s,\quad
(\s\da_{z_n}\s+\s{_1\over^\kappa}\s H_n\s)\s\theta\s=\s0\s,\
\s n=1,\dots,N\s.
\label{KZ}
\qqq
Above, $H_n\equiv H_n(\tau,\un{z})\s$ are operators acting on the
CS states and $\kappa$ is a coupling constant.
The KZB equations may be interpreted as
the horizontality equations
for a KZB connection in bundle $\CW$.
The consistency of the equations (the flatness of the
KZB connection) requires that the operators $H_n,
\ n\geq0$, commute for fixed $(\tau,\un{z})$.
In fact, $H_n,\ n\geq 1,$ are quantum versions of the
quadratic classical Hamiltonians $h_n$ of the
elliptic Hitchin system which Poisson-commute.
In the case of elliptic curve with no insertions
the conformal blocks coincide with the (linear combinations of)
characters of the integrable representations
of the affine algebras. $H_0$ in eq.\s\s(\ref{KZ})
is then proportional to the Laplacian on the Cartan
algebra and the KZB equations reduce to the well
known heat equation for the elliptic theta functions.
In the case of elliptic curve with one insertion,
$H_0(\tau)$ becomes a version of the Calogero-Sutherland
Hamiltonian \cite{Etin-1}. For many insertions, operators
$H_n(\tau,\un{z}),\ n\geq 1$, are elliptic versions
of the Gaudin Hamiltonians \cite{Gaud}.
\vskip 0.2cm

A hermitian structure on a holomorphic vector bundle
induces a unique unitary connection of the 1,0 type.
One expects \cite{Karp} that the scalar product of
the CS states induces this way the KZB connection.
For the elliptic case with no insertions, that
was shown implicitly in \cite{GK} where it was proven
that the affine characters form an orthonormal basis
of the space of CS states. For group $SU(2)$
in the elliptic case with one insertion the unitarity
of the KZB connection w.r.t. the scalar product
of the CS states was proven in \cite{FG}. In the present
paper we study this problem for the general elliptic situation.
We show that, assuming convergence of the integrals
giving the scalar product, the unitarity of the elliptic
KZB connection follows from a result announced in
\cite{FV} which provides a basis for the Bethe-Ansatz
diagonalization of the commuting operators
$H_n(\tau,\un{z}),\ n\geq0$. The relation of the integral
representations for the genus zero conformal blocks of the WZW
theory to the Bethe Ansatz was observed in \cite{Bab}, see also
\cite{BabFl,RV,FFR}. The elliptic $SU(2)$ case counterparts
of these relations go back to XIXth century works
of Hermite on the Lam\'{e} operator, as noticed in \cite{Etin-1}.
Recently, the integral representations for the elliptic
conformal blocks were used in \cite{FV} to obtain the Bethe-Ansatz
treatment of the elliptic Calogero-Sutherland model,
an open problem till then. Our work exhibits an intrinsic
connection between the Bethe-Ansatz and the unitarity
of the KZB connection.
\vskip 0.2cm

The paper is organized as follows. In Sect.\s\s 2, we
describe briefly the space of CS states in the holomorphic
quantization and identify the states on the elliptic
curve as vector-valued theta-functions. Sect.\s\s 3
is devoted to the scalar product of the elliptic
CS states. As realized in \cite{GK} for the case with no
insertions, the formal functional integral over the gauge
fields giving the scalar product may be computed as an
iterative Gaussian integral. The insertions for general
group $G$ are handled by combining the methods of
refs.\s\s\cite{GK,FGK}. In Sect.\s\s4, we describe
the KZB connection and recall its relations to the Hitchin
integrable systems. Sect.\s\s5 discusses the unitarity
of the KZB connection and finally, in Sect.\s\s6,
we consider the relations to the Bethe Ansatz.
\vskip 0.9cm

\nsection{Elliptic CS states}
\vskip 0.5cm

Let us recall the description of the CS states
on the elliptic curve $\CT_\tau=\NC/(\NZ+\tau\NZ)$
with insertion points $z_n$, given in \cite{fg,FG},
with due modifications
required by the replacement of the group $SU(2)$
by $G$. For simplicity, we assume $G$ to be simple, connected
and simply connected and will denote by $\Ng$ its
Lie algebra. Let $\CA$ denote the (complex) vector space
of the $0,1$-components $A\equiv A_{\bar z}d\bar z$ of
the $\Ng$-valued gauge fields. The group $\CG$ of
complex gauge transformations $g:\CT_\tau\rightarrow G^\NC$
acts on $\CA$ by
\qq
A\ \mapsto\ {}^g\hspace{-1mm}A=gAg^{-1}+g\de g^{-1},
\label{cgt}
\qqq
with $\de=d\bar z\partial_{\bar z}$.
The Chern-Simons states $\Psi$ are holomorphic
functionals on $\CA$. For the case with insertions
of time-like Wilson lines, they take values in the tensor
product $\otimes_n V_{\la_n}\equiv\un{V}$
of the irreducible representation spaces of $G$ of highest
weights $\la_n$, associated to the insertions.
States $\Psi$ verify the chiral Ward identity
\qq
\Psi(^g\hskip -1mm A)=\ee^{kS(g^{-1},A)}
\otimes_n g(z_n)_{_{(n)}}\Psi(A)\label{chwi}
\qqq
where $S(g,A)$ is the action of the
Wess-Zumino-Witten model coupled to
$A$, see \cite{fg}, and the subsubscript
$(n)$ indicates that the group element acts
in the factor $V_{\la_n}$ of the tensor product space
$\un{V}$.
\vs 0.3cm

Restricting functionals $\Psi$ to connections
$A_u=\pi\s u d\bar z/\tau_2$
for $u$ in the Cartan algebra $\Nh^\NC$
and $\tau_2\equiv{\rm Im}\s\tau$,
we can assign to every state a holomorphic map
$\gamma:\Nh^ \NC\rightarrow\un{V}$
related to $\Psi$ by the equation
\qq
\Psi(A_u)= \ee^{\pi k |u|^2/(2\tau_2)}
\otimes_n (\ee^{\pi(z_n-\bar z_n)u/\tau_2})_{_{(n)}}
\gamma(u)\label{gamma}
\qqq
where $|u|^2\equiv\lle u,u\rle$ with the Killing form
$\lle\s\cdot\s,\s\cdot\s\rle$ normalized so that $|\alpha|^2=2$
for long roots $\alpha$ (we identify $\Ng$ with its dual).
Holomorphic maps $\gamma$
corresponding to states $\Psi$ satisfy the following conditions
\qq
\gamma(u+q^\vee)\ &=&\gamma(u)\quad{\rm for}\ q^\vee\ {\rm
in\ the\ coroot\ lattice}\ Q^\vee,\alabel{conda}a\cr\cr
\gamma(u+\tau q^\vee)&=&\ee^{-\pi ik\lle q^\vee\hs{-0.04cm},\s
\tau q^\vee+2u\rle}
\otimes_n(\ee^{-2\pi iz_n q^\vee})_{_{(n)}}\gamma(u),\alabel{condb}b\cr\cr
0&=&\oplus_n(h)_{(n)}\gamma(u)\quad{\rm for}\ h\in\Nh,\alabel{condc}c\cr\cr
\gamma(wuw^{-1}))&=&\otimes_n (w)_n\gamma(u)\quad{\rm for}
\ w\in G\ {\rm normalizing}\ \Nh,\alabel{condd}d\cr
\qqq
\vskip -0.4cm
\noindent and
\vskip -0.8cm
\qq
\left(\sum\limits_n\ee^{-2\pi i sz_n}(e_\alpha)_{_{(n)}}\right)^p
\gamma(u+tp^\vee)\s=\s\CO(t^p)
\label{regu}
\qqq
for any root $\alpha$, coweight $p^\vee$ satisfying
$\lle p^\vee,\alpha\rle=1$, $u$ s.t. $\lle u,\alpha\rle=m+\tau s$
with $m,s$ integers, $p=1,2,...$ and $t\to 0$. Here and below
$e_\alpha$ is the step generator of ${\bf g}^\NC$ corresponding
to root $\alpha$.
\vskip 0.2cm

We shall denote the space of maps $\gamma(u)$
satisfying the above conditions by $W_{\tau,\un z,\un\la}$.
Conditions (\ref{conda},a) and (\ref{condb},b) mean that
$\gamma$ is a vector of theta-functions which
due to condition (\ref{condc},c) take
values in the zero weight subspace ${\un{V}}_0\subset\un{V}$
and which by condition (\ref{condd},d) are Weyl group
covariant. Equations (\ref{regu}) specify their regularity
on the hyperplanes $\lle u,\alpha\rle\in\NZ+\tau\NZ$ where
$\alpha\in\Nh$ are roots of the algebra $\Ng$. They have been
obtained for $G=SU(2)$ in \cite{fg} and for general $G$
in \cite{Etin}, see also \cite{FW}. They follow
by demanding the regularity at $t=0$ of the maps
$$t\s\mapsto\s\Psi(^{h_{ms}}\hs{-0.12cm}
A_{u'\hspace{-0.05cm},t})=\Psi(^{g_{ms}}
\hs{-0.13cm}A_{u+tp^\vee})\s\s$$
where $\s t\mapsto A_{u'\hspace{-0.05cm},t}={\pi\over\tau_2}
(u'+tp^\vee+e_\alpha)\s d\bar z\s$,
\s with $u'=u-(m+\tau s)p^\vee$, is a 1-parameter holomorphic family
of gauge fields, $\s h_{ms}(z)=\exp[{_{\pi}
\over^{\tau_2}}((m+\bar\tau s)z
-(m+\tau s)\bar z)p^\vee]\s$ is a multivalued gauge transformation
and $\s g_{ms}(z)=h_{ms}(z)\exp[-t^{-1}e_\alpha]\s h_{ms}^{-1}(z)$
is a univalued one. Since $\lle u'\hspace{-0.05cm},
\alpha\rle=0$, \s$A_{u'+tp^\vee}$
may be gauge transformed to $A_{u'\hspace{-0.05cm},t}$
for $t\not=0$ by a constant
gauge transformation $\exp[t^{-1}e_\alpha]$).
Gauge fields ${}^{h_{ms}}\hspace{-0.12cm}A_{u'\hspace{-0.05cm},0}$
lie on codimension one strata in the space of gauge fields
which cannot be attained by gauge transforming $A_u$'s.
Conditions (\ref{regu}) assure the global regularity of $\Psi$
but, due to the properties (\ref{conda},a,b), they are not
independent for different $m$ and $s$. For example, for all simple
groups different from $SU(2)$ or $SO(2r+1)$ it is enough to take
$m=s=0$.
\vs 0.9cm

\nsection{Scalar product of CS states}
\vs 0.5cm

The scalar product of Chern-Simons states is formally
given by the functional integral
\qq
\parallel \Psi\parallel^2=
\int \vert\Psi(A)\vert^2\s\s
\ee^{-{ik\over 2\pi}\int\tr(A^* A)}
\ DA\s DA^*\label{sp}
\qqq
We shall perform the above functional integration
reducing the scalar product expression to a
finite-dimensional integral which, if finite,
will provide $W_{\tau,\un z,\un\la}$
with a natural structure of a Hilbert space.
The finiteness has been proven in the special
cases (general $G$ with no insertions \cite{GK}
or $SU(2)$ with one insertion \cite{FG}) and has
been conjectured to hold in general \cite{Karp}.
The strategy for the calculation of the integral (\ref{sp})
will be as in \cite{GK} and \cite{FG} so we shall be brief
discussing in detail only the treatment of the insertions
essentially borrowed from \cite{FGK}.
\vskip 0.5cm

\subsection{Change of variables}
\vskip 0.3cm

\noindent We shall reparametrize the gauge fields
in the functional integral (\ref{sp}) as
\qq
A\s=\s^{g^{-1}}\hskip-1.5mm A_u
\label{rep}
\qqq
with $g$ a complex gauge transformation and
$u$ in some fundamental domain
of the action of translations
$u\mapsto u+q^\vee$,
$u\mapsto u+\tau q^\vee$
on $\Nh^\NC$. \s(\s$\Nh^\NC/(Q^\vee+\tau Q^\vee)/{\rm Weyl\ group}$
is the space of gauge orbits of semistable gauge fields; ignoring
the Weyl group action produces an overall factor
in the scalar product equal to the order of the Weyl group.)
We shall use the Iwasawa decomposition of the $G^\NC$-valued
field $g$\s:
\qq
g\s=\s \exp[\sum\limits_{\alpha>0}
v_\alpha\s e_\alpha]\s\exp[\phi/2]\s U\s\equiv\s n\s\exp[\phi/2]
\s U\s\equiv b\s U
\label{trian}
\qqq
where $\phi$ takes values in
the Cartan algebra $\Nh$ and $U$ in the compact group $G$.
Upon the change of variables, field $U$ decouples from
the functional integral (\ref{sp}) due to the gauge invariance
leaving us with a WZW-type functional integral
over the $G^\NC/G$-valued fields $b$ \cite{GK}
\qq
\parallel \Psi\parallel^2&
=&\int \langle\Psi(A_u)\s,\s
\otimes_n (b b^*)_{(n)}^{-1}(z_n)\s\Psi(A_u)\rangle
\cr
\ &&\hspace{0.4cm}
\times\s\s\ee^{kS(bb^*,\s A_u+A_u^{\s*})\s-\s{ik\over 2\pi}
\int\tr(A_u^* A_u)}\ j(u,b)\ \delta(\phi(0))\ Db\ d^2u\s
\label{newvar}
\qqq
where $Db=\prod_z db(z)$ is the formal product of $G^\NC$-invariant
measures on $G^\NC/G$, the delta function fixes the remaining
freedom in the parametrization (\ref{rep}) and the
Jacobian of the change of variables
\qq\non
j(u,b)\s=\s{\rm const}.\s\s\tau_2^{-2r}\s\s\ee^{\s2h^\vee
S(bb^*,\s A_u+A_u^{\s*})}\ \det(\de_{A_u}^{\s\dagger}\de_{A_u})\s.
\qqq
Above, $r$ stands for the rank of $G$, $h^\vee$ for its dual
Coxeter number and $\de_{A_u}\equiv\de+[A_u,\s\cdot\s\hs{0.03cm}]$.
The last determinant was computed in \cite{GK}:
\qq\non
\det(\de_{A_u}^{\s\dagger}\de_{A_u})\s=\s
{\rm const}.\s\s\tau_2^{2r}\s\ee^{\s \pi h^\vee|u-u^*|^2/\tau_2}
\s|\Pi(u)|^4
\qqq
where $\Pi$ is the Kac-Weyl denominator:
\qq\non
\Pi(u)\s=\s \ee^{\s \pi i d\tau/12}
\prod\limits_{\alpha>0}(\ee^{\s\pi i\lle u,\alpha\rle}-
\ee^{-\pi i\lle u,\alpha\rle})
\prod\limits_{l>0}(1-q^l)^r\prod\limits_{\alpha}(1-q^l\ee^{\s2\pi
i\lle u,\alpha\rle})
\qqq
with $d$ denoting the dimension of the group and
$q\equiv\ee^{2\pi i\tau}$.
The WZW action in the parametrization (\ref{trian})
takes the form
\qq
S(bb^*,A_u+A_u^{\s*})\s=\s
-{i\over 4\pi}\int\lle\da\phi,\de\phi
\rle-{i\over 2\pi}\int\lle\ee^{\phi_u}(n_u^{-1}\de n_u)^*\ee^{-\phi_u}
\hs{-0.04cm},\s n_u^{-1}\de n_u\rle
\nonumber\qqq
where
\qq\non
n_u\s=\s\ee^{-\pi(z-\bar z)u/\tau_2}n\s\ee^{\s\pi(z-\bar z)u/\tau_2}
\s,\quad\phi_u\s=\s\phi-\pi(z-\bar z)(u-\bar u)/\tau_2\s.
\qqq
Note that
\qq\non
n_u\s=\s\exp[\sum\limits_{\alpha>0}\ee^{-\pi\lle u,\alpha\rle
(z-\bar z)/\tau_2}\s v_\alpha\s e_\alpha\s]
\s\equiv\s\exp[\sum\limits_{\alpha>0}v'_\alpha\s e_\alpha\s]\s.
\qqq
In terms of the Iwasawa variables, the invariant
measure on $G^\NC/G$ is
\qq\non
db\s=\s \prod\limits_{j=1}^rd\phi^j\prod\limits_{\alpha>0}
d^2(\ee^{-\lle\phi,\alpha\rle/2}v_\alpha)
\qqq
where $\phi^j=\lle h_j,\phi\rle$ are the coordinates
of $\phi$ w.r.t. an orthonormal basis $(h_j)$
of $\Nh$. Using the holomorphic functions $\ga(u)$
to represent $\Psi$ and the parametrization (\ref{trian}),
we obtain
\qq\non
\lle\s\Psi(A_u)\s,\s\prod_n(bb^*)^{-1}_{_{(n)}}(z_n)\s\Psi(A_u)\s\rle
\s=\s\ee^{{\pi k\over\tau_2}\s{\rm Re}(|u|^2)}\s
\lle\s\ga(u)\s,\otimes_n((n_u^*)^{-1}\ee^{-\phi_u}
n_u^{-1})_{_{(n)}}(z_n)\s\ga(u)\s\rle\s.\hs{0.3cm}
\qqq
Finally,
\qq\non
-{_{ik}\over^{2\pi}}\smallint\tr(A_u^* A_u)\s
=\s-\pi k\lle\bar u,u\rle/\tau_2\s.
\qqq
With all these ingredients,
\qq
\parallel \Psi\parallel^2&=&{\rm const.}\s\int
\lle\s\ga(u)\s,
\otimes_n((n_u^*)^{-1}\ee^{-\phi_u}
n_u^{-1})_{_{(n)}}(z_n)\s\ga(u)\s\rle
\s\s\s\exp[-{_{i(k+2h^\vee)}\over^{4\pi}}
\smallint\lle\da\phi,\de\phi\rle]\cr
&&\times\s\s\exp[-\s{_{i(k+2h^\vee)}\over^{2\pi}}
\smallint\lle\ee^{\phi_u}(n_u^{-1}\de n_u)^*\ee^{-\phi_u}
\hs{-0.04cm},\s n_u^{-1}\de n_u\rle
\s+\s{_{\pi(k+2h^\vee)}\over^{2\tau_2}}|u-\bar u|^2]\cr
&&\times\s\s|\Pi(u)|^4\s\s\delta(\phi(0))\s\s
\s d^{2r}u\s\prod\limits_{j=1}^r D\phi^j
\prod\limits_{\alpha>0,\s z}
d^2(\ee^{-\lle\phi_u(z),\alpha\rle/2}v'_\alpha(z))\s.
\label{nfi}
\qqq
\vskip 0.1cm

In order to render the $n_u$-dependent terms in the action
of the last functional integral quadratic, we shall
introduce new variables $(\eta'_\alpha)_{\alpha>0}$ defined by
\qq
n_u^{-1}\de n_u\s=\s\sum\limits_{\alpha>0}\de\eta'_\alpha e_\alpha\s.
\label{chv}
\qqq
Since $n_u(z+1)=n_u(z)$ and $n_u(z+\tau)=\ee^{-2\pi i u}n_u(z)
\ee^{\s2\pi i u}$, it follows that
\qq
\eta'_\alpha(z+1)=\eta'_\alpha(z)\quad
\eta'_\alpha(z+\tau)=\ee^{-2\pi i\lle u,\alpha\rle}
\eta'_\alpha(z)
\label{pert}
\qqq
and that the change of variables is well defined since
$\de$ is invertible (for generic $u$) on functions satisfying
the periodicity conditions (\ref{pert}). It is easy to see
that the change of variables $(v'_\alpha)\s\mapsto\s(\eta'_\alpha)$
has a triangular nature: $\eta'_\alpha=v'_\alpha + F_\alpha
((v'_\beta)_{\beta<\alpha})$
where $\beta<\alpha$ if $\alpha-\beta$ is a positive root.
Hence, the formal volume element does not change:
\qq\non
\prod\limits_{\alpha>0,\s z}
d^2(\ee^{-\lle\phi_u(z),\alpha\rle/2}v'_\alpha(z))\s
=\s\prod\limits_{\alpha>0,\s z}
d^2(\ee^{-\lle\phi_u(z),\alpha\rle/2}\eta'_\alpha(z))\s.
\qqq
Although the action in the functional integral (\ref{nfi}),
which was polynomial in variables $v'_\alpha$
(and their derivatives) becomes quadratic in (derivatives of)
$\eta'_\alpha$, there persists the $n_u$-dependence of
the insertions in the functional integral (\ref{nfi}).
Thus, we have to invert relation (\ref{chv}) in order to
express $n_u$ as a function of $\eta'\equiv\sum\limits_{\alpha>0}
\eta_\alpha' e_{\alpha}$. This will be done generalizing
a trick of \cite{FGK}.
\vskip 0.2cm

Let $\Nn=\bigoplus\limits_{\alpha>0}\NC e_{\alpha}$ be the
nilpotent subalgebra of $\Ng^\NC$. The corresponding group
$N\subset G^\NC$ may be mapped into the enveloping
algebra $\CU(\Nn)$ (no completion problem arises if we
work in highest weight irreducible representations of $G^\NC$).
$\CU(\Nn)$ is graded by the positive cone $Q_+$ in the root
lattice $Q$:
\qq\non
\CU(\Nn)\s=\s\bigoplus\limits_{q\in Q_+}\CU(\Nn)_q
\qqq
according to the eigenvalues of the adjoint action
of $\Nh$, with $e_{\alpha_1}\cdots e_{\alpha_m}\in\CU(\Nn)_{\alpha_1
+\dots+\alpha_m}$. The map $z\to n_u^{-1}(z)\equiv\nu(z)$
may be viewed as taking values in $\CU(\Nn)$ and it satisfies
then the twisted periodicity conditions
\qq
\nu_q(z+1)=\nu_q(z),\quad \nu_q(z+\tau)
=\ee^{-2\pi i\lle u,q\rle}\nu_q(z)
\label{twi}
\qqq
where $\nu_q$ denotes the $\CU(\Nn)_q$-component of $\nu$.
Relation (\ref{chv}) may be now rewritten as an equation
\qq\non
\da_{\bar z} n_u^{-1}\s=\s -(\da_{\bar z}\eta')\s n_u^{-1}\s,
\qqq
for $\CU(\Nn)$-valued functions and solved with use of
the Green functions of the twisted $\da_{\bar z}$-operator.
Since $(n_u^{-1})_0=1$, we obtain
\qq
n_u^{-1}\s=\s\sum\limits_{K=0}^\infty (-1)^K
\s\s\smash{\mathop{\da_{\bar z}^{-1}\s(\da_{\bar z}\eta')\s\cdots
\s\da_{\bar z}^{-1}\s(\da_{\bar z}\eta')}_{K\ {\rm times}}}
\cdot 1\s.
\label{nu}
\qqq
The Green function of $\da_{\bar z}$ acting
on functions obeying conditions (\ref{twi})
may be easily expressed by the Jacobi
theta function $\vartheta_1(z)\s=\s\sum_l(-1)^l
\ee^{\s\pi i(l+1/2)^2\tau+2\pi i z(l+1/2)}$ satisfying
\qq
\vartheta_1(z+1)=-\vartheta_1(z),\quad\vartheta_1(z+\tau)=
-\ee^{-\pi i(\tau+2z)}\vartheta_1(z)\s.
\label{theta1}
\qqq
Explicitly, it is equal to
\qq
P_x(z)\s=\s{{\vartheta'_1(0)\s\vartheta_1(x
+z)}\over{\pi\s\vartheta_1(x)\s\vartheta_1(z)}}
\label{P}
\qqq
with $x={\lle u,q\rle}$. Hence eq.\s\s(\ref{nu}) takes the form
\qq
n_u^{-1}(z)\s=\s\sum\limits_{K=0}^\infty(-1)^K
\hs{-0.4cm}\sum\limits_{(\alpha_1,\dots,\alpha_K)}
\hspace{-0.4cm}\int P_{\lle u,\alpha_1
+\dots+\alpha_K\rle}(z-y_1)\s(\da_{\bar z}\eta'_{\alpha_1})(y_1)\s
P_{\lle u,\alpha_2+\dots+\alpha_K\rle}(y_1-y_2)\cr
\times\s\s(\da_{\bar z}\eta'_{\alpha_2})(y_2)\
\cdots\cdots\ P_{\lle u,\alpha_K\rle}(y_{K-1}-y_K)\s
(\da_{\bar z}\eta'_{\alpha_K})(y_K)\s\s e_{\alpha_1}
\dots e_{\alpha_K}\s\s d^2y_1\cdots d^2y_K\s.\hspace{0.5cm}
\label{nu1}
\qqq
In the $\ga(u)$ matrix element, we may insert the partition
of unity $\otimes_n(\sum\limits_{a_n}|\mu_{a_n},a_n\rle
\lle\mu_{a_n},a_n|)$ where vectors $|\mu_{a_n},a_n\rle$
corresponding to weights $\mu_{a_n}$ form an orthonormal
basis in the representation spaces $V_{\la_n}$. This gives
\qq\non
\lle\s\ga(u)\s,
\otimes_n((n_u^*)^{-1}\ee^{-\phi_u}
n_u^{-1})_{_{(n)}}(z_n)\s\ga(u)\s\rle&&\cr\cr
&&\hs{-5cm}\s=\s\sum\limits_{\un{a}}\prod\limits_n
\ee^{-\lle\phi_u(z_n),
\mu_{a_n}\rle}\ |\otimes_n\lle\mu_{a_n},a_n|\s\otimes_n
n_u^{-1}(z_n)\s\gamma(u)\s\rle|^2
\qqq
into which we may insert the expressions (\ref{nu1}) for
$n_u^{-1}(z_n)$. As in \cite{FGK}, the final renormalization
of the functional integral over the Cartan algebra-valued
field $\phi$ will kill most of the terms obtained this way
leaving only the ones with $|\mu_{a_n},a_n\rle$ equal
to the highest weight vectors $|\la_n\rle$ and the sequences
of positive roots $(\alpha_1,\dots,\alpha_p)$ composed
uniquely of simple roots. We may then write
\qq
\lle\s\ga(u)\s,
\otimes_n((n_u^*)^{-1}\ee^{-\phi_u}
n_u^{-1})_{_{(n)}}(z_n)\s\ga(u)\s\rle
\s=\s\prod\limits_n\ee^{-\lle\phi_u(z_n),\la_n\rle}\s\s
\bigg\vert\sum\limits_{\un{K}}\sum\limits_{\un\alpha}\int
F_{\un{K}_,\un{\alpha}}(\tau,u,\un{z},\un{y})\cr
\times\s\s\prod\limits_n\prod\limits_{i=1}^{K_n}(\da_{\bar z}
\eta'_{\alpha_{n,i}})(y_{n,i})\s\s d^2y_{n,i}
\ \lle\la\s|\otimes_n(e_{\alpha_{n,1}}\cdots
e_{\alpha_{n,K_n}})\s\gamma(u)\s\rle\s\bigg\vert^2\ \
+\ \ .\ \ .\ \ .\ \ \s
\hspace{0.5cm}
\label{me}
\qqq
where $\lle\la|\s\equiv\otimes_n\lle\la_n|\s$,
$\s\un{\alpha}=(\alpha_{1,1},\dots,\alpha_{1,K_1},\alpha_{2,1},
\dots\dots,\alpha_{N,K_N})\s\equiv\s(\alpha_1,\dots,\alpha_K)\s$
is a sequence of $K=\sum_nK_n$ simple roots satisfying
\qq
\displaystyle{\sum\limits_{s=1}^K\alpha_s=\sum\limits_{n=1}^N\la_n\s,}
\label{R}
\qqq
$\s\un{y}=(y_{1,1},\dots,y_{1,K_1},y_{2,1},
\dots\dots,y_{N,K_N})\s\equiv\s(y_1,\dots,y_K)\s$
is a sequence of $K$ points in $\CT_\tau$ and
the kernels $F_{{\un p},\un{\alpha}}$ are composed of the
Green functions of twisted $\de$
\qq
F_{\un{K},\un{\alpha}}(\tau,u,\un{z},\un{y})\s=\s
\prod\limits_n
P_{\lle u,\alpha_{n,1}+\dots+\alpha_{n,K_n}\rle}
(z_n-y_{n,1})\s
P_{\lle u,\alpha_{n,2}+\dots+\alpha_{n,K_n}\rle}
(y_{n,1}-y_{n,2})\cr
\times\ \cdots\
P_{\lle u,\alpha_{n,K_n}\rle}(y_{n,K_n-1}-y_{n,K_n})\s.
\label{F}
\qqq
"\ .\ \ .\ \ .\ " \s contains the terms that will drop under
the renormalization of the $\phi$-integral.
\vskip 0.5cm

\subsection{Functional integration}
\vskip 0.3cm

The above use of eq.(\ref{nu}) to express the $\eta'$ dependence
of the insertions reduces the $\eta'$ integral to the form
\qq
\int\hs{-0.1cm}\prod\limits_{s=1}^K
(\da_{\bar z}\eta'_{\alpha_s})(y_s)
\s\overline{(\da_{\bar z}\eta'_{\beta_s})(v_s)}
\ \exp[\s{_{-i(k+2h^\vee)}\over^{2\pi}}\sum\limits_{\alpha>0}
\smallint\ee^{-\lle\phi_u,\alpha\rle}\overline{\de\eta'_\alpha}
\s\de\eta'_\alpha\s]\prod\limits_{\alpha>0,\s z}
d^2(\ee^{-\lle\phi_u,\alpha\rle/2}\eta'_\alpha)\cr
=\s({_\pi\over^{k+2h^\vee}})^K
\sum\limits_{\sigma\in S_K}\prod\limits_{s=1}^K
\ee^{\s\lle\phi_u(y_s),\alpha_s\rle}\s\s
\delta_{\alpha_s,\beta_{\sigma(s)}}\s\s\delta(y_s-v_{\sigma(s)})\s
\prod\limits_{\alpha>0}\det(\de_\alpha^{\s\dagger}\de_\alpha)^{-1}\s\s
\hs{0.5cm}
\label{wick}
\qqq
with $\sigma$ running over the permutations of $K$ points
and with $\de_\alpha\equiv \ee^{-\lle\phi_u,\alpha\rle/2}\de\s
\ee^{\s\lle\phi_u,\alpha\rle/2}\s$. The determinants are
well known
\qq\non
\prod\limits_{\alpha>0}\det(\de_\alpha^{\s\dagger}
\de_\alpha)^{-1}=\s{\rm const}.\s\s
\ee^{-\pi r\tau_2/6}\s|\Pi(u)|^{-2}
\prod\limits_{l=1}^\infty|1-q^l|^{2r}\s\cr
\times\s\s\exp[\s{_{ih^\vee}
\over^{4\pi}}\smallint\lle\da\phi,\de\phi\rle-{_{\pi h^\vee}
\over^{2\tau_2}}|u-\bar u|^2]\s.
\qqq
After the $\eta'_\alpha$-integration
and an easy combinatorial manipulation, see \cite{FGK},
trading the sum over root sequences $\un{\alpha}$ into sums
over permutations (two $\un\alpha$'s satisfying eq.\s\s(\ref{R})
differ necessarily only by a permutation),
the scalar product formula (\ref{nfi}) becomes
\qq
\parallel \Psi\parallel^2&=&{\rm const.}\s\s
\ee^{-\pi r\tau_2/6}\prod\limits_{l=1}^\infty|1-q^l|^{2r}
\int\prod\limits_{n=1}^N\ee^{-\lle\phi_u(z_n),\s\la_n\rle}
\prod\limits_{s=1}^K\ee^{\s\lle\phi_u(y_s),\s\alpha_s\rle}\s\cr
&&\hs{-1.6cm}\times\s\s\exp[\s-{_{i\kappa}\over^{4\pi}}
\smallint\lle\da\phi,\de\phi\rle\s+\s{_{\pi\kappa}
\over^{2\tau_2}}|u-\bar u|^2\s]\s\s\s\delta(\phi(0))\s\s
\prod\limits_{j=1}^r D\phi^j
\ |\Pi(u)|^{2}\s\cr
&&\hs{-1.6cm}\times\s\bigg|\sum\limits_{\un{K}}
\sum\limits_{\sigma\in S_K}F_{\un{K},\sigma\un{\alpha}}
(\tau,u,\un{z},\sigma\un{y})
\s\s\lle\la|\otimes_n(e_{(\sigma\alpha)_{n,1}}
\cdots\s e_{(\sigma\alpha)_{n,K_n}})_{_{(n)}}\s\ga(u)\s\rle\bigg|^2
\s d^{2r}u\s\prod\limits_{s=1}^Kd^2y_s\s\s
\label{nnfi}
\qqq
where $\kappa\equiv k+h^\vee$ and $\un{\alpha}$ is a fixed sequence
of $K$ simple roots satisfying (\ref{R}).
\vskip 0.3cm

The remaining $\phi$-integral is of the Gaussian (Coulomb gas) type
and may be easily performed:
\qq
&\int\ee^{\int\sum_j\phi^j(y) f^j(y)}\ \exp[\s-{_{i\kappa}\over^{4\pi}}
\smallint\lle\da\phi,\de\phi\rle\s]\ \delta(\phi(0))\s
\prod\limits_{j=1}^r D\phi^j&\cr
&=\s{\rm const}.\ \tau_2^{r/2}\s\s\det'(\de^\dagger\de)^{-r/2}\
\ee^{-{1\over{2(k^+h^\vee)}}\sum_j\int f^j(y)\CG(y-y')f^j(y')}&\cr\cr
&=\s{\rm const}.\ \tau_2^{-r/2}\s\ee^{\pi r\tau_2/6}
\prod\limits_{l=1}^\infty|1-q^l|^{-2r}\
\ee^{-{1\over{2(k^+h^\vee)}}\sum_j\int f^j(y)\CG(y-y')f^j(y')}&
\label{phint}
\qqq
where we should take $\s f=-\sum_n\la_n\delta_{z_n}
+\sum_s\alpha_s\delta_{y_s}\s$ which corresponds in the Coulomb gas
jargon to external charges at points $z_n$ and screening
charges at points $y_s$. The Green function is
\qq
\CG(z)\s=\s\ln|\vartheta_1(z)|^2+{_{\pi}\over^{2\tau_2}}(z-\bar z)^2+
{\rm const}.
\label{Gree}
\qqq
Renormalizing the divergences due to the singularities
at coinciding points by point-splitting, as explained in \cite{FGK}
and \cite{FG}, we finally obtain for the scalar product of CS states
the following finite-dimensional integral expression:
\qq
\parallel \Psi\parallel^2\s=\s{\rm const.}\s\s\tau_2^{-r/2}
\int\ee^{\s{_{\pi \kappa}
\over^{2\tau_2}}|w-\bar w|^2}\left\vert\s
\ee^{-{1\over\kappa}S(\tau,\un{z},\un{y})}
\s\s\lle\s G(\tau,u,\un{z},\un{y})\s,\s\theta(u)\s\rle
\s\right\vert^2
\s d^{2r}u\prod\limits_{s=1}^Kd^2y_s\s\s\
\label{scpr}
\qqq
where $\theta(u)\s\equiv\s\Pi(u)\s\gamma(u)$,
\qq
w\s\equiv\s u+{_1\over^{\kappa}}\sum\limits_{n=1}^Nz_n\la_n
-{_1\over^{\kappa}}\sum\limits_{s=1}^Ky_s\alpha_s\s,
\label{w}
\qqq
$S$ is a multivalued holomorphic functions of $\tau$
and non-coincident $z_n$'s and $y_s$'s and
$G$ a holomorphic multivalued map with values in the dual
space $\un{V}^*$ of $\un{V}$ given, respectively, by
\qq
S(\tau,\un{z},\un{y})\s=\s
\sum\limits_{n<n'}\lle\la_n,\la_{n'}\rle
\ln\tilde\vartheta_1(z_n-z_{n'})&-&
\sum\limits_{n,s}\lle\la_n,\alpha_s\rle
\ln\tilde\vartheta_1(z_n-y_s)\hs{2.3cm}\cr
&&\hs{1cm}+\s\sum\limits_{s<s'}\lle\alpha_s,
\alpha_{s'}\rle\ln\tilde\vartheta_1(y_s-y_{s'})
\label{S}
\qqq
\vskip -0.4cm
\noindent with $\s\tilde\vartheta_1(z)
\equiv\vartheta_1(z)/\vartheta'_1(0)\s$
and
\qq
G(\tau,u,\un{z},\un{y})\s=\s\sum\limits_{\un K}
\sum\limits_{\sigma\in S_K}F_{\un{K},\sigma\un{\alpha}}
(\tau,u,\un{z},\sigma\un{y})
\s\s\lle\la|\otimes_n(e_{(\sigma\alpha)_{n,1}}
\cdots e_{(\sigma\alpha)_{n,K_n}})_{_{(n)}}\s.
\label{G}
\qqq
Using the transformation properties (\ref{theta1}),
it is straightforward to verify that for
\s$H\equiv
\ee^{-{1\over\kappa}S}\s
\lle G,\theta\rle\s$, \s$\un{\delta}^n=
(0,\dots\smash{\mathop{1}\limits_{\hat n}}\dots,0)$
and $\un{\delta}^s$ defined similarly,
\qq
&&\hbox to 3.1cm{$H(\tau,u+q^\vee,\un{z},\un{y})$\hfill}
=\ H(\tau,u,\un{z},\un{y})\s,\cr
&&\hbox to 3.1cm{$H(\tau,u+\tau q^\vee,\un{z},\un{y})$\hfill}=\
\ee^{-\pi i\kappa\lle q^\vee,\s\tau q^\vee+2w\rle}\s\s
H(\tau,u,\un{z},\un{y})\s,\cr
&&\hbox to 3.1cm{$H(\tau,u,\un{z}+\un{\delta}^n,\un{y})
$\hfill}
=\ (-1)^{\lle\la_n,\la_n\rle/\kappa}\s\s
H(\tau,u,\un{z},\un{y})\s,\clabel{H}\cr
&&\hbox to 3.1cm{$H(\tau,u,\un{z}+\tau\un{\delta}^n,\un{y})
$\hfill}
=\ (-1)^{\lle\la_n,\la_n\rle/\kappa}\s\s\ee^{-\pi i\tau\lle\la_n,
\la_n\rle/\kappa\s+\s2\pi i\lle w,\la_n\rle}\s\s
H(\tau,u,\un{z},\un{y})\s,\cr
&&\hbox to 3.1cm{$H(\tau,u,\un{z},\un{y}+\un{\delta}^s)$\hfill}
=\ (-1)^{\lle\alpha_s,\alpha_s\rle/\kappa}\s\s
H(\tau,u,\un{z},\un{y})\s,\cr
&&\hbox to 3.1cm{$H(\tau,u,\un{z},\un{y}+\tau\un{\delta}^s)$\hfill}
=\ (-1)^{\lle\alpha_s,\alpha_s\rle/\kappa}
\s\s\ee^{-\pi i\tau\lle\alpha_s,
\alpha_s\rle/\kappa\s-\s2\pi i\lle w,\alpha_s\rle}\s\s
H(\tau,u,\un{z},\un{y})\s.
\qqq
It is then easy to see that the under-integral expression
in (\ref{scpr}) is univalued under the $u\mapsto u+(\tau)q^\vee$,
$z_n\mapsto z_n+(\tau)1$, $y_s\mapsto y_s+(\tau)1$ transformations.
\vskip 0.3cm

It is useful to compare the above expression for the scalar
product of the CS states with the genus zero ones obtained
in \cite{FGK}. The genus zero states are determined by
their values $\Psi(0)\equiv\gamma$ at $A=0$ belonging
to the $G$-invariant
subspace $\un{V}^G$ of the the tensor product $\un{V}$ of
the representation spaces. Adapting the notations to
the ones of the present paper, eqs.\s\s(19.20)
of \cite{FGK} read
\qq
\parallel \Psi\parallel^2&=&{\rm const.}\s
\int\left\vert\s\ee^{-{1\over\kappa}S^0(\un{z},\un{y})}
\s\s \lle\s G^0(\un{z},\un{y})\s,\s\gamma\s\rle\s\right\vert^2
\s\prod\limits_{s=1}^Kd^2y_s
\label{scpr0}
\qqq
where
\qq
S^0(\un{z},\un{y})\s=\s
\sum\limits_{n<n'}\lle\la_n,\la_{n'}\rle
\ln(z_n-z_{n'})&-&\sum\limits_{n,s}\lle\la_n,\alpha_s\rle
\ln(z_n-y_s)\hs{1.5cm}\cr
&&\hs{1.8cm}+\s\sum\limits_{s<s'}
\lle\alpha_s,\alpha_{s'}\rle
\ln(y_s-y_{s'})
\label{S0}
\qqq
\vskip -0.4cm
\noindent and
\qq
G^0(\un{z},\un{y})\s=\s\sum\limits_{\un k}
\sum\limits_{\sigma\in S_K}F_{\un{K}}
(\un{z},\sigma\un{y})
\s\s\lle\la|\otimes_n(e_{(\sigma\alpha)_{n,1}}
\cdots\s e_{(\sigma\alpha)_{n,K_n}})_{_{(n)}}
\label{G0}
\qqq
with
\qq
F_{\un{K}}(\un{z},\un{y})\s=\s
{_1\over^{\pi^{K}}}\prod\limits_n
{_1\over^{z_n-y_{n,1}}}\s
{_1\over^{y_{n,1}-y_{n,2}}}\s\cdots\s
{_1\over^{y_{n,K_n-1}-y_{n,K_n}}}\s.
\label{F0}
\qqq
The similarity to the genus one case is obvious.
\vskip 0.9cm

\nsection{Knizhnik-Zamolodchikov(-Bernard) connection and
the\break Hitchin systems}
\vskip 0.5cm

The trivial bundle with the fiber $\un V$ over the space
$X^0_N\equiv\NC^N\setminus\Delta$
where $\Delta$ contains vectors $\un z$
with coinciding components carries a 1-parameter family
of flat holomorphic connections defined by
\qq\non
\nabla_{\bar z_n}\s=\s\da_{\bar z_n}\s,\quad
\nabla_{z_n}\s=\s\da_{z_n}\s+\s {_1\over^\kappa}\s H^0_n(\un z)\s\s
\qqq
where $H^0_n(\un{z})$ are the Gaudin Hamiltonians \cite{Gaud}:
\qq
H^0_n(\un z)\s=\s\sum\limits_{a=1}^d
\sum\limits_{n'\not=n}{t^a_{_{\s(n)}}t^a_{_{\s(n')}}\over
z_{n'}-z_{n}}
\label{Gaud}
\qqq
with $(t^a)$ forming an orthonormal basis of Lie algebra $\Ng$.
Connection $\nabla$ appeared (implicitly) for the first time
in ref.\s\s\cite{KZ}: it was shown there that the
genus zero conformal blocks $\gamma$ of the WZW theory satisfy
the (Knizhnik-Zamolodchikov) equations $\nabla\gamma=0$.
In fact, the WZW conformal blocks are horizontal sections
of a (generally proper) subbundle
$\CW^0\subset X^0_N\times\un{V}^G$. The fibers $W^0_{\un{z}}$
of $\CW^0$ may be identified by the assignment $\Psi\to\Psi(0)$
with the genus zero CS state spaces. The subbundle
$\CW^0$ may be described by
giving explicit algebraic conditions, depending holomorphically
on $\un{z}$, on the invariant tensors in $\un{V}^G$ \cite{TK,jaWZW}.
The KZ connection $\nabla$ preserves the subbundles
$X_N^0\times\un{V}^G$ and $\CW^0$ of the trivial
bundle $X_N^0\times\un{V}$.
\vskip 0.3cm

The extension of the Knizhnik-Zamolodchikov
connection to the genus one case was first obtained
in ref.\s\s\cite{B1} and elaborated further
in \cite{fg,Etin-1,FW,FV}. We shall use the description of
the genus one CS states by the theta functions
$\theta(u)=\Pi(u)\s\ga(u)$. The spaces
$W_{\tau,\un{z}}$ of states form a holomorphic bundle
over the space of pairs $(\tau,\un{z})$ with
no coincidences in $z_n$'s viewed as points in
$\CT_\tau$. The holomorphic sections of $\CW$
correspond to holomorphic families $(\tau,\un{z})\rightarrow
\theta(\tau,\un{z},\s\cdot\s)$.
The KZB connection is given by the formulae
\qq
\nabla_{\bar\tau}\s=\s\da_{\bar\tau},\ \quad
\nabla_{\bar z_n}\s=\s\da_{\bar z_n},\ \quad
\nabla_{\tau}\s=\s\da_{\tau}\s+\s{_1\over^\kappa}\s
H_0(\tau,\un{z})\s,\ \quad
\nabla_{z_n}=\da_{z_n}\s+\s{_1\over^\kappa}\s
H_n(\tau,\un{z})\s,
\label{KZB}
\qqq
compare to eqs.\s\s(\ref{KZ}). Explicitly \cite{Etin-1,FW,FV},
\qq
H_0(\tau,\un{z})&=&{_i\over^{4\pi}}\Delta_u\s+\s
\sum\limits_{n,n'=1}^N\bigg(
{_i\over^4}\sum\limits_\alpha \da_x
P_{\lle u,\alpha\rle}(z_{n}-z_{n'})
\s\s(e_\alpha)_{_{(n)}}(e_{-\alpha})_{_{(n')}}\cr
&&\hspace{3cm}+\s{_i\over^{8\pi}}
\sum\limits_{j=1}^r(\rho(z_{n}-z_{n'})^2+
\rho'(z_{n}-z_{n'}))\s\s h^j_{_{(n)}}h^j_{_{(n')}}
\bigg)\s,\label{H0}\\
H_n(\tau,\un{z})&=&-\sum\limits_{j=1}^rh^j_{_{(n)}}\da_{u^j}\s-
\s\sum\limits_{n'\not=n}\bigg(\sum\limits_{\alpha}
\pi\s P_{\lle u,\alpha\rle}(z_{n}-z_{n'})\s\s
(e_\alpha)_{_{(n)}}(e_{-\alpha})_{_{(n')}}\cr
&&\hspace{6.09cm}+\s\sum\limits_{j=1}^r
\rho(z_{n}-z_{n'}))\s\s h^j_{_{(n)}}h^j_{_{(n')}}
\bigg)
\label{Hn}
\qqq
where $\Delta_u\equiv\sum_j(\da_{u^j})^2$
and $\rho\equiv\vartheta'_1/\vartheta_1$.
The expressions for $H_n,\ n\geq 1,$ reduce to the ones
for the Gaudin Hamiltonians (\ref{Gaud}) in the limit
$\tau\to i\infty$.
The operators $H_n,\s n\geq0,$ acting, say, on
meromorphic functions of $u\in\Nh^\NC$ with values
in $\un{V}_0$ commute, see \cite{FW}. Their commutation
forms part of the conditions assuring the flatness of the KZB
connection (\ref{KZB})\hs{0.03cm}. Although the coefficient
functions in $H_n$ have poles on the hyperplanes
$\lle u,\alpha\rle\in\NZ+\tau\NZ$, the connection maps
holomorphic families of genus one CS states
into families with the same property, due
to the increased regularity (\ref{regu}) of the CS
states on the singular hyperplanes, see \cite{fg,FW}.
\vskip 0.3cm

The commuting operators $H^0_n$ for genus zero
or $H_n$ at genus one are quantizations of classical
Poisson-commuting Hamiltonians of the, respectively,
genus zero and genus one Hitchin integrable
systems \cite{Hitch,DonWitt,Nekr,ER}.
Let us briefly recall this relation. Given a Riemann surface
$\Sigma$ and a group $G$, let $\CA$ denote the corresponding space
of $0,1$-gauge fields and $\CG$ the group of complex gauge
transformations, as in the beginning of Sect.\s\s 2. Both
are (infinite dimensional) complex manifolds and we shall
work in the holomorphic category. The cotangent bundle
$T^*\CA$ is composed of pairs $(A,\Phi)$ where
$\Phi$ is a $1,0$ form on $\Sigma$ with values in Lie
algebra $\Ng^\NC$. The action (\ref{cgt})
of $\CG$ on $\CA$ lifts to the symplectic action
$\s(A,\Phi)\s\mapsto\s({}^g\hspace{-1mm}A,\s{}^g\hspace{-0.4mm}
\Phi\equiv g\Phi g^{-1})\s$ on $T^*\CA$ with the moment map
\qq\non
M(A,\Phi)\s=\s\de A+A\wedge\Phi+\Phi\wedge A
\qqq
(the 2-forms on $\Sigma$ with values in $\Ng^\NC$ form
the space dual to the Lie algebra of $\CG$).
If $\un{z}$ is a sequence of $N$ insertion points in
$\Sigma$ and $\un{\la}$ a corresponding sequence
of Cartan algebra elements, then we may symplectically
reduce $T^*\CA$ w.r.t.\s\s the $\CG$-coadjoint orbit of
$\sum_n\la_n\delta_{z_n}$ defining the reduced phase space
\qq\non
\CP_{\un{z},\un{\la}}\s=\s M^{-1}(\{\sum_n\la_n\delta_{z_n}\})
\bigg/\CG_{\un{z},\un{\la}}
\qqq
where $\CG_{\un{z},\un{\la}}$ is the subgroup of
$\CG$ fixing $\sum_n\la_n\delta_{z_n}$.
The symplectic form on $T^*\CA$ descends to
(the non-singular part of) $\CP_{\un{z},\un{\la}}$
turning it into a finite dimensional (complex) symplectic
manifold (we shall ignore here the singularities of
$\CP_{\un{z},\un{\la}}$). For a $G^\NC$
invariant homogeneous polynomial $p$ on $\Ng^\NC$
of degree $d_p$, the assignment
\qq\non
(A,\Phi)\s\longmapsto\s p(\Phi)
\qqq
defines a map on $T^*\CA$ with values in the space
of $d_p$-differentials (sections of the $d_p$-th
symmetric power of the canonical bundle on $\Sigma$).
All those vector-valued maps Poisson-commute since they
depend only
on $\Phi$. They descend to $\CP_{\un{z},\un{\la}}$
giving a system of Poisson-commuting maps $h_p$ with values
in the (finite-dimensional) spaces of meromorphic
$d_p$-differentials with poles of order $\leq d_p$
at the insertion points $z_n$. Their components form a maximal
set of classical Hamiltonians in involution turning
$\CP_{\un{z},\un{\la}}$ into an integrable system
introduced and analyzed in \cite{Hitch} for the case
without insertions, see \cite{Nekr,ER} for the generalization
including the insertions.
\vskip 0.2cm

Let us specify first the above construction to the
genus zero case. In that case, (almost) each gauge
field $A$ is in the gauge orbit of $A=0$, i.e.\s\s it is
of the form
\qq\non
A=h^{-1}\de h
\qqq
for $h\in\CG$ with $h$ determined
modulo $h\mapsto h_0h$ with constant $h_0$.
Equation $\s M(A,\Phi)=\sum_n\la_n\delta_{z_n}\s$
becomes now
\qq\non
\de({}^h\hspace{-0.4mm}\Phi)\s=\s\sum_nh(z_n)\la_n h^{-1}(z_n)
\s\s\delta_{z_n}
\qqq
which has a unique solution
\qq\non
({}^h\hspace{-0.4mm}\Phi)(z)\s=\s{_1\over^{2\pi i}}
\sum_n{{h(z_n)\la_nh^{-1}(z_n)}\over{z-z_n}}\s dz
\qqq
provided that the sum of the residues vanishes:
\qq\non
\sum_nh(z_n)\la_nh^{-1}(z_n)\s=\s0\s.
\qqq
The group $\CG_{\un z,\un{\la}}$ is composed of
arbitrary gauge transformations $g$ s.t. $g(z_n)\in G^\NC_{\la_n}$,
acting on $h$ by $h\mapsto hg^{-1}$. Above, $G^\NC_{\la_n}$
denotes the subgroups of the $G^\NC$ stabilizing
$\la_n$ under the adjoint action.
For the symplectically reduced phase space
$P^0_{\un{z},\un{\la}}$
(the superscript $0$ referring to the genus zero case)
we obtain
\qq\non
P^0_{\un{z},\un{\la}}\ \cong\
\{\s\s\un{\mu}\s\s\s|\s\s\s\mu_n\in
\CO_{\la_n}\s,\ \sum_n\mu_n=0\s\s\}\bigg/ G^\NC
\qqq
where $\mu_n\equiv h(z_n)\la_n h^{-1}(z_n)$ run
through the (co)adjoint orbits $\CO_{\la_n}$.
The latter are naturally complex symplectic manifolds
and it is not difficult to check that, as a symplectic
manifold, $P^0_{\un{z},\un{\la}}$ is the reduction
of $\smash{\mathop{\times}\limits_{^n}}\CO_{\la_n}$
by the diagonal action of
$G^\NC$. Since for an invariant polynomial $p$, $\s p(\Phi)=
p({}^h\hskip -0.4mm\Phi)$, \s the corresponding
Poisson-commuting Hamiltonians
on $P^0_{\un{z},\un{\la}}$ are
\qq\non
h^0_p(\un{\mu})(z)\s=\s p\left({_1\over^{2\pi i}}
\sum_n{{\mu_n}
\over{z-z_n}}\right)(dz)^{d_p}\s.
\qqq
In particular, for the quadratic
polynomial given by the Killing form
$p_2=\lle\s\cdot\s,\s\cdot\s\rle$, we obtain
the quadratic meromorphic differential
\qq\non
h^0_{p_2}(\un{\mu})(z)\s&=&\s -{_1\over^{4\pi^2}}\sum_{n,n'}
{{\lle\mu_n,\mu_{n'}\rle}\over{(z-z_n)(z-z_{n'})}}\s\s(dz)^2\cr
&=&
\sum_n \left(-{\lle\mu_n,\mu_{n}\rle\over 4\pi^2(z-z_n)^2}+
{1\over z-z_n}h^0_n\right)(dz)^2
\qqq
where the residues at $z=z_n$ are
\qq\non
h^0_n\s={_1\over^{2\pi^2}} \sum_{n'\not=n}{{\lle\mu_n,\mu_{n'}\rle}
\over{z_{n'}-z_n}}.
\qqq
These are, up to normalization, the classical versions of the
Gaudin Hamiltonians of eq.\s\s(\ref{Gaud}). The latter
may be obtained from $h^0_n$'s by replacing the coordinates
$\mu^a_n\equiv\lle\mu_n,t^a\rle$ on the coadjoint orbit
$\CO_{\la_n}$ by the generators $t^a_{_{(n)}}$ of $\Ng$ acting
in the irreducible representation $V_{\la_n}$ obtained by
geometric quantization of $\CO_{\la_n}$.
\vskip 0.4cm

Similarly at genus one, for $\Sigma=\CT_\tau$, (almost) each gauge
field is in the gauge orbit of the gauge fields $A_u$,
i.e.\s\s it is of the form
\qq\non
A\s=\s {}^{h^{-1}}\hskip -1mm A_u\s=\s (h_uh)^{-1}\de(h_uh)
\qqq
with $u\in\Nh^\NC$ and $h_u\equiv\ee^{\pi\bar z(u-\bar u)/\tau_2}$.
Hence the gauge fields $A$ may be parametrized
by pairs $(u,h)$ with the identifications
\qq
(u,h)\s=\s(wuw^{-1},\s wh)\s=\s(u+q^\vee,\s h_{q^\vee}^{-1}h)\s=\s
(u+\tau q^\vee,\s h_{\tau q^\vee}^{\s-1}h)
\label{ident}
\qqq
for $q^\vee$ in the coroot lattice $Q^\vee$ and $w$ in
the normalizer $N(\Nh^\NC)\subset G^\NC$ of $\Nh^\NC$.
Equation $\s M(A,\Phi)=\sum_n\la_n\delta_{z_n}\s$
becomes now
\qq
\de\s{}^{(h_uh)}\hskip -0.5mm \Phi\s
=\s\sum_n(h_uh)(z_n)\la_n(h_uh)^{-1}(z_n)\s\s\delta_{z_n}\s.
\label{shs}
\qqq
Upon decomposing
\qq\non
(h_uh)(z_n)\la_n(h_uh)^{-1}(z_n)\ =\
\sum_\alpha\mu_n^{-\alpha}\s e_{\alpha}\s\s+\s\s\mu_n^0\ \s
\equiv\ \s\mu_n
\qqq
with $\mu_n^0$ in the Cartan subalgebra $\Nh^\NC$,
eq.\s\s(\ref{shs}) may be solved, provided that
$\sum_n\mu_n^0=0$, with use of the Green functions
of the twisted and untwisted $\de$-operator on $\CT_\tau$:
\qq\non
{}^{(h_uh)}\hskip -0.4mm \Phi(z)\s=\s
\Phi^0 dz\s+\s
{_1\over^{2\pi i}}
\sum_n\left(\pi\sum_\alpha P_{\lle u,\alpha\rle}(z-z_n)
\s\mu_n^{-\alpha}\s e_\alpha\s+\s\rho(z-z_n)\s\mu_n^0\right) dz
\qqq
where $\Phi^0$ is an arbitrary constant in $\Nh^\NC$.
The symplectically reduced phase space $P_{\un{z},\un{\la}}$
becomes
\qq\non
P_{\un{z},\un{\la}}\ =\ \{\s\s(u,\Phi^0,\un{\mu})\in T^*\Nh^\NC\times
(\smash{\mathop{\times}\limits_{^n}}\CO_{\la_n})\s\s\s\vert\s\s\s
\sum_n\mu_n^0=0\s\s\}\bigg/N(\Nh^\NC)\s\s{\dl o}\s
\s(Q^\vee+\tau Q^\vee)
\qqq
with the identifications
\qq\non
(u,\Phi^0,\un{\mu})=(wuw^{-1},\s w\Phi^0 w^{-1},
\s w\un{\mu}w^{-1})=
(u+(\tau)q^\vee,\Phi^0,\s(h_{(\tau)q^\vee}^{-1}(z_n)\mu_n
h_{(\tau)q^\vee}(z_n)))\s.\ \
\qqq
As a symplectic manifold, $P_{\un{z},\un{\la}}$ is a symplectic
reduction of $T^*\Nh^\NC\times(
\smash{\mathop{\times}\limits_{^n}}\CO_{\la_n})$ by
the group $N(\Nh^\NC)\s\s{\dl o}\s\s(Q^\vee+\tau Q^\vee)$.
The Hitchin Hamiltonians
become
\qq\non
h_p(u,\Phi^0,\un{\mu})(z)&&\cr\cr
&&\hs{-2cm}\s=\s
p\bigg(\Phi^0\s+\s{_1\over^{2\pi i}}
\sum_n\bigg(\pi\sum_\alpha P_{\lle u,\alpha\rle}(z-z_n)
\s\mu_n^{-\alpha}\s e_\alpha\ +\ \rho(z-z_n)\s\mu_n^0\bigg)
\bigg) (dz)^{d_p}
\qqq
which for $p=p_2$ and upon writing ${h}_{p_2}(u,\Phi^0,\un{\mu})(z)=
{\tilde h}_{p_2}(u,\Phi^0,\un{\mu})(z)\s(dz)^2$, reduces to
\qq\non
{\tilde h}_{p_2}(u,\Phi^0,\un{\mu})(z)\s=\s
\lle \Phi^0,\s\Phi^0\rle\s
+\s{_1\over^{\pi i}}
\sum_n\rho(z-z_n)\s\lle\mu^0_n,\s\Phi^0\rle\s
-\s{_1\over^{4\pi^2}}\sum_{n,n'}\bigg(
\sum_\alpha\pi^2\cr
\times\s\s P_{\lle u,\alpha\rle}(z-z_n)\s
P_{-\lle u,\alpha\rle}(z-z_{n'})\s\s\mu^{-\alpha}_n\s
\mu^{\alpha}_{n'}\s+\s\rho(z-z_n)\s\rho(z-z_{n'})\s\lle\mu^0_n,
\s\mu^0_{n'}\rle\bigg)\hskip-1cm \cr\cr =\s\s
{_1\over^{4\pi^2}}\sum_n \rho'(z-z_n) \lle \mu_n,\mu_n\rle
+\sum_n \rho(z-z_n)\s h_n(u, \Phi^0, \un\mu) + h_0 (u,\phi^0,\un\mu).
\qqq
with the residues at $z_n$
\qq\non
h_n(u,\Phi^0,\un{\mu})\s=\s{_1\over^{\pi i}}
\s\lle\mu^0_n,\s\Phi^0\rle\s-
\s{_1\over^{2\pi^2}}\sum_{n'\not=n}\bigg(
\sum_\alpha\pi\s P_{\lle u,\alpha\rle}
(z_{n}-z_{n'})\s\s\mu^\alpha_n\s
\mu^{-\alpha}_{n'}\hs{1cm}\cr
+\s\s\rho(z_{n}-z_{n'})\s\s\lle\mu^0_n,
\s\mu^0_{n'}\rle\bigg)
\qqq
and the holomorphic (actually $z$ independent) piece
\qq\non
h_0(u,\phi^0,\un\mu)&=&\lle \Phi^0,\Phi^0 \rle -
{_1\over^{8\pi^2}}\sum_{n,n'}\bigg( 2\pi\sum_\alpha
\partial_x P_{\lle u,\alpha\rle}(z_n-z_{n'})
\mu_n^\alpha\mu_{n'}^{-\alpha}\cr
&&\hskip 3.5cm  +\s\s(\rho(z_n-z_{n'})^2
+\rho'(z_n-z_{n'}))\s\lle \mu^0_n,\mu^0_n \rle\bigg).
\qqq
Hamiltonians $h_n$, $n\geq 0$ are the classical versions
of the elliptic Gaudin
Hamiltonians $H_n$ of eqs.\s\s(\ref{H0}) and \pref{Hn},
see \cite{Nekr}\cite{Olsha}.
\vskip 0.3cm

The $SU(n)$ elliptic Hitchin system corresponding to one
insertion has unexpectedly appeared recently in the
description of the low energy sector of supersymmetric
Yang-Mills theories \cite{DonWitt}.
\vskip 0.9cm

\nsection{Unitarity of the KZB connection}
\vskip 0.5cm

One of the essential features of the structures
discussed above should be the compatibility of
the KZ and KZB connections with the scalar product of
CS states. The integrals in eq.\s\s(\ref{scpr0}) have been conjectured
in \cite{Quadr} (and proved in many cases) to converge precisely
for invariant tensors $\gamma\in W^0_{\un z}$ and to equip
bundle $\CW^0$ with the hermitian structure preserved
by the KZ connection. The latter condition means that
for all local holomorphic sections $\un{z}\mapsto\gamma(\un{z})$
of bundles $\CW^0$, corresponding to holomorphic families
$\un{z}\mapsto\Psi(\un{z})$ of CS states,
\qq
\da_{z_n}\parallel \Psi\parallel^2\s=\s
(\s\Psi\s,\s\nabla_{z_n}\Psi\s)\s.
\label{uni0}
\qqq
In order to see why one should expect such a relation,
it will be convenient to express the scalar product
integral in the language of differential forms, following
refs.\s\s\cite{VarSch,VarSch1}.
Let $\omega(y)\equiv{1\over\pi}\s y^{-1}dy$.
Introduce the $\un{V}^*$-valued $K$ forms
\qq\non
\Omega^0_{\un{K},\un{\alpha}}(\un{z},\un{y})&=&
\omega(z_1-y_{1,1})\wedge\omega(y_{1,1}-y_{1,2})\wedge\s\dots\wedge
\omega(y_{1,K_1-1}-y_{1,K_1})\cr
&&\wedge\ \ .\ \ .\ \ .\ \ .\ \ .\ \ .\ \ .\ \ .\ \ .\ \ .\ \ .
\ \ .\ \ .\ \ .\ \ .\ \ .\ \ .\ \ .\ \ .\ \ .\ \ .\ \ .
\ \ .\ \ .\ \ .\cr
&&\wedge\omega(z_N-y_{N,1})\wedge\omega(y_{N,1}-y_{N,2})\wedge\s\dots\wedge
\omega(y_{N,K_N-1}-y_{N,K_N})\cr
&&\hs{5.7cm}\times\s\s\lle\la|\otimes_n(e_{\alpha_{n,1}}
\dots{}^{^{\s\s}}e_{\alpha_{n,K_n}})_{_{(n)}}
\qqq
\vskip -0.3cm
\noindent and
\vskip -0.3cm
\qq\non
\Omega^0(\un{z},\un{y})\s=\s\sum\limits_{\un{K}}
\sum\limits_{\sigma\in S_K}(-1)^{|\sigma|}\s
\Omega_{\un{K},\sigma\un{\alpha}}(\un{z},\sigma\un{y})\s.
\qqq
We may rewrite the scalar product formula (\ref{scpr0}) as
\qq\non
\parallel \Psi\parallel^2\s=\s{\rm const.}\s
\int\limits_{Y_{\un{z}}}
\left\vert\s\ee^{-{1\over\kappa}S^0}
\s\s\lle\s\Omega^0\s,\s\gamma\s\rle\s\right\vert^2\s\s
\qqq
where $Y_{\un{z}}$ stands for the space of $\un{y}$'s with
$y_s$'s not coinciding among themselves and with
$z_n$'s. We use the conventions that $|\Omega^0|^2\equiv\s
(-1)^{K(K-1)/2}({_i\over^2})^K\s\Omega^0\wedge
\overline{\Omega^0}$ and that the integral of the forms of degree
lower than the dimension of the cycle is zero. Assuming a sufficiently
strong convergence of the integrals, we may enter with the
holomorphic exterior derivative under the integral so that
\qq\non
\da\parallel \Psi\parallel^2\s
=\s{\rm const.}\s
\int\limits_{Y_{\un{z}}}
\da\left\vert\s\ee^{-{1\over\kappa}S^0}
\s\s\lle\s\Omega^0\s,\s\gamma\s\rle\s\right\vert^2&&\cr
&&\hspace{-3.6cm}=\s{\rm const.}\s
\int\limits_{Y_{\un{z}}}
\da\hs{-0.05cm}\left(
\ee^{-{1\over\kappa}S^0}
\s\s\lle\s\Omega^0\s,\s\gamma\s\rle\right)\wedge
\s\overline{\ee^{-{1\over\kappa}S^0}
\s\s\lle\s\Omega^0\s,\s\gamma\s\rle}\s\s
\qqq
where the $\da$-operator on the l.h.s. acts on the $z$-variables
and the one under the integral may be taken as
acting on both $z$- and $y$-variables. The $\un{V}^*$-valued
holomorphic multivalued forms $\Omega^0$
have two basic properties:
\qq
\da\s\Omega^0\s=\s 0\quad\quad{\rm and}\quad\quad
(\da S^0)\wedge\Omega^0\s+\s\sum\limits_{n=1}^Ndz_n\wedge H^0_n\s
\Omega^0\s=\s0
\label{btr}
\qqq
with the contragradient action of the Gaudin Hamiltonians on the
$\un{V}^*$-valued form $\Omega^0$. The first relation is trivial.
The second, more involved one, has been proven in \cite{VarSch1}.
Using eqs.\s\s(\ref{btr}), we obtain
\qq\non
\da\parallel \Psi\parallel^2\s
=\s{\rm const.}\s
\int\limits_{Y_{\un{z}}}
\left\vert\s\ee^{-{1\over\kappa}S^0}\s\right\vert^2
\s\s\sum_ndz_n\wedge\lle\s\Omega^0\s,\s
(\da_{z_n}+{_1\over^\kappa}H_n)\s\gamma
\s\rle\s\wedge\s\overline{\ee^{-{1\over\kappa}S^0}
\s\s\lle\s\Omega^0\s,\s\gamma\s\rle}
\qqq
which implies the relation (\ref{uni0}).
\vskip 0.4cm

The above analysis has its counterpart for the elliptic
case. We again conjecture that the integrals in
eq.\s\s(\ref{scpr}) converge for $\theta$ corresponding
to CS states (in fact exactly when the regularity
conditions (\ref{regu}) are fulfilled) and that
the resulting hermitian structure on the bundle
$\CW$ renders the KZB connection unitary, see \cite{FG}
for the proof of this conjecture for the case of $G=SU(2)$
and one insertion.
In order to substantiate the conjecture for the case
of general $G$ and arbitrary insertions,
let us rewrite the genus one scalar product
integral (\ref{scpr}) in the language of differential
forms. Define
\qq\non
\omega_{q}(z)\equiv P_{\lle u,q\rle}(z)\s dz
\s-\s{_i\over^{2\pi}}\s\da_xP_{\lle u,q\rle}(z)\s d\tau\s,
\qqq
see eq.\s\s(\ref{P}). It is easy to see that $\da\s\omega_q=0$,
if $\da$ differentiates only the variables $(\tau,z)$.
Set
\qq
\Omega_{\un{K},\un{\alpha}}(\tau,u,\un{z},\un{y})&=&
\omega_{\beta_{1,1}}(z_1-y_{1,1})
\wedge\omega_{\beta_{1,2}}
(y_{1,1}-y_{1,2})\wedge\s\dots\wedge
\omega_{\beta_{1,K_1}}(y_{1,K_1-1}-y_{1,K_1})\hs{0.75cm}\cr
&&\hs{-1.4cm}\wedge\ \ .\ \ .\ \ .
\ \ .\ \ .\ \ .\ \ .\ \ .\ \ .\ \ .\ \ .
\ \ .\ \ .\ \ .\ \ .\ \ .\ \ .\ \ .\ \ .\ \ .\ \ .\ \ .
\ \ .\ \ .\ \ .\ \ .\ \ .\ \ .\ \ .\ \ .\ \ .\ \ .\ \ .\cr
&&\hs{-1.4cm}\wedge\omega_{\beta_{N,1}}(z_N-y_{N,1})\wedge
\omega_{\beta_{N,2}}(y_{N,1}-y_{N,2})\wedge\s\dots\wedge
\omega_{\beta_{N,K_{N_{\s}}}}\hs{-0.02cm}(y_{N,K_N-1}-y_{N,K_N})\cr
&&\hs{6.2cm}\times{}^{^{\s\s}}\s
\lle\la|\otimes_n(e_{\alpha_{n,1}}
\dots{}^{^{\s\s}}e_{\alpha_{n,K_n}})_{_{(n)}}
\label{Omeg}
\qqq
with $\s\beta_{n,i}=\sum\limits_{i'=i}^{K_n}\alpha_{n,i'}\s$ and
\qq\non
\Omega(\tau,u,\un{z},\un{y})\s=\s
\sum\limits_{\un{K}}\sum\limits_{\sigma\in S_K}(-1)^{|\sigma|}\s
\Omega_{\un{K},\sigma\un{\alpha}}(\tau,u,\un{z},\sigma\un{y})\s.
\qqq
We still have to dress $\Omega$ with the $du^j$ differentials.
The convenient way to do this is to define the form
\qq
{\NOm}\s=\s \ee^{-{1\over\kappa}S}\bigg(
dw^1\wedge\dots\wedge dw^r
\wedge\lle\Omega,\theta\rle\s+
\s{_1\over^{2\pi i\kappa}}\sum\limits_{j=1}^r
dw^1\wedge\dots\smash{\mathop{d\tau}\limits_{\hat j}}\dots
\wedge dw^r\s\da_{u^j}\lle\Omega,\theta\rle\bigg)\hs{0.4cm}
\label{dress}
\qqq
where $w^j\equiv\lle h^j,w\rle$ with $w$ given by
eq.\s\s(\ref{w}) and, as before, $\theta\equiv\Pi\ga$.
It is easy to see that eq.\s\s(\ref{scpr}) may be rewritten as
\qq
&&\parallel \Psi\parallel^2\s=\s
{\rm const.}\s\s\tau_2^{-r/2}
\int\limits_{UY_{\tau,\un{z}}}
\ee^{\s{_{\pi \kappa}
\over^{2\tau_2}}|w-\bar w|^2}\left\vert\s
\ee^{-{1\over\kappa}S}
\s\s\lle\s\Omega\s,\s\theta\s\rle\s\right\vert^2
\s d^{2r}u\cr
&&\hspace{6.5cm}\s=\s{\rm const.}\s\s\tau_2^{-r/2}
\int\limits_{UY_{\tau,\un{z}}}
\ee^{\s{_{\pi \kappa}
\over^{2\tau_2}}|w-\bar w|^2}\left\vert\s
\s\NOm\s\right\vert^2\s.
\label{dble}
\qqq
This holds since the terms with $d\tau$ differentials
do not contribute to the
integral over the cycle $UY_{\tau,\un{z}}$ composed of
points $(u,\un{y})$ with identifications
$u=u+(\tau)q^\vee$,
$y_s=y_s+(\tau)1$
which form the $(r+K)$-dimensional
torus $\CT_\tau^{r+K}$ (with coincidences of $y_s$'s among themselves
and with $z_n$'s removed). The gain from adding the terms with
$d\tau$'s is that the form $\NOm$ transforms
under the maps $u\mapsto u+(\tau)q^\vee$, $y_s\mapsto y_s+(\tau)1$
(and $z_n\mapsto z_n+(\tau)1$) exactly as the functions $H$, see
eq.\s\s(\ref{H}), and, as the result, the form under the last integral
in eq.\s\s(\ref{dble}) is well defined on $UY_{\tau,\un{z}}$.
\vskip 0.2cm

We would like to compute the holomorphic differential
of the norms squared $\parallel\Psi\parallel^2$ of
a holomorphic family of states. Due to the fact that the
integration cycles $UY_{\tau,\un{z}}$ depend nontrivially
on the differentiation variables, this requires a more
subtle consideration than in the genus zero case where
we could ignore this problem (assuming enough convergence
at the coinciding points). Ignoring again the coinciding
points, the geometric setup is as follows.
We are given a bundle $\varpi:\CN\rightarrow \CM$ of
multidimensional tori over
the space $\CM$ of pairs $(\tau,\un{z})$. Let us forget for
a moment the complex structures on these spaces considering
them as real manifolds. On $\CN$ we are given a differential
form $\eta$ of degree equal to the dimension of the fiber
(we shall not need an obvious generalization of the story
to forms of arbitrary degree) and we are studying a function
$f$ on the base obtained by fiber-wise integration of $\eta$
\qq\non
f(n)\s=\s\int\limits_{\varpi^{-1}(\{n\})}\eta\s\s.
\qqq
In local trivialization of the bundle, only the components
of $\eta$ with differentials along the fiber contribute
to $f$. If there are no convergence problems, then obviously
\qq
df(n)\s=\s\int\limits_{\varpi^{-1}(\{n\})}d^\perp\eta
\s=\s\int\limits_{\varpi^{-1}(\{n\})}d\eta
\label{uint}
\qqq
where the differential $d^\perp$ in the transverse directions
is defined using a trivialization and the second equality
follows since $d^\parallel\eta\equiv
(d-d^\perp)\eta$ is a closed form when restricted to the
fibers so that its fiber-wise integral vanishes (again assuming
no convergence problems).
\vskip 0.3cm

Let us first see how this argument may be used
in the simplest case with no insertions where
we have
\qq\non
\parallel \Psi\parallel^2\s=\s
{\rm const.}\s\s\tau_2^{-r/2}
\int\limits_{\CT_\tau^{\s r}}
\ee^{\s{_{\pi \kappa}
\over^{2\tau_2}}|u-\bar u|^2}\left\vert
\theta(u)\right\vert^2\s d^{2r}u\s=\s
{\rm const.}\s\s\tau_2^{-r/2}
\int\limits_{\CT_\tau^{\s r}}
\ee^{\s{_{\pi \kappa}
\over^{2\tau_2}}|u-\bar u|^2}\left\vert
\NOm\right\vert^2
\qqq
where $\theta(u+q^\vee)=\theta(u)$ and $\theta(u+\tau q^\vee)
=\ee^{-\pi i\kappa\lle q^\vee\hs{-0.04cm},\s
\tau q^\vee\hs{-0.04cm}+2u\rle}\s\theta(u)$,
i.\s e.\s\s$\theta$ is an $r$-dimensional
theta-function of degree $2\kappa$, and
\qq\non
\NOm\s=\s\theta\s\s du^1\wedge\dots\wedge du^r
\s+\s{_1\over^{2\pi i\kappa}}\sum_{j=1}^r\da_{u^j}\theta\s
\s du^1\wedge\dots\smash{\mathop{d\tau}\limits_{\hat j}
}\dots\wedge du^r
\qqq
is a holomorphic $r$-form which under $u\mapsto u+(\tau)q^\vee$
transforms the same way as the theta-function $\theta$.
Note the relations:
\qq
du^j\wedge\NOm\s=\s{_i\over^{2\pi\kappa}}\s d\tau\wedge\da_{u^j}
\NOm\s,\quad\quad\sum_jdu^j\wedge\da_{u^j}\NOm\s=\s{_i\over
^{2\pi\kappa}}\s d\tau\wedge\Delta_u\NOm\s.
\label{rella}
\qqq
The $(2r)$-form $\s\eta\s=\s\ee^{\s{_{\pi \kappa}
\over^{2\tau_2}}|u-\bar u|^2}\left\vert\NOm\right\vert^2\s$
is well defined on the bundle $\CN$ of tori obtained by varying
$\CT_\tau^{\s r}$ with $\tau$. Applying the above geometrical
considerations to the case at hand, we obtain
\qq\non
d\parallel \Psi\parallel^2\s=\s
{\rm const.}\s\s\tau_2^{-r/2}
\int\limits_{\CT_\tau^{\s r}}
(d+{_{ri}\over^{4\tau_2}}(d\tau-d\bar\tau)\wedge)\left(
\ee^{\s{_{\pi \kappa}
\over^{2\tau_2}}|u-\bar u|^2}\left\vert
\NOm\right\vert^2\right)\s.
\qqq
Clearly, only the holomorphic exterior derivative $\da$ will
contribute under the integral to \s$\da\parallel
\Psi\parallel^2\s$. Explicit differentiation gives:
\qq
\da\parallel \Psi\parallel^2&=&
{\rm const.}\s\s\tau_2^{-r/2}
\int\limits_{\CT_\tau^{\s r}}
\ee^{\s{_{\pi \kappa}
\over^{2\tau_2}}|u-\bar u|^2}\bigg(\bigg(
{_{\pi i\kappa}\over^{4\tau_2^2}}|u-\bar u|^2\s
d\tau\wedge\s+\s{_{\pi\kappa}\over^{\tau_2}}
\sum_j(u^j-\bar u^j)\s du^j\wedge\s\hs{0.5cm}\cr
&&\hs{4.2cm}+\s{_{ri}\over^{4\tau_2}}\s d\tau\wedge\s
+\s d\tau\wedge\da_\tau\s+\s\sum_j du^j\wedge\da_{u^j}
\bigg)\NOm\bigg)\wedge
\s\overline{\NOm}\s\hs{0.5cm}\cr
&=&{\rm const.}\s\s\tau_2^{-r/2}
\int\limits_{\CT_\tau^{\s r}}
\ee^{\s{_{\pi \kappa}
\over^{2\tau_2}}|u-\bar u|^2}\bigg(\bigg(
{_{\pi i\kappa}\over^{4\tau_2^2}}|u-\bar u|^2\s
d\tau\wedge\s+\s{_{i}\over^{2\tau_2}}
\s d\tau\wedge\sum_j(u^j-\bar u^j)\s\da_{u^j}\s\hs{0.5cm}\cr
&&\hs{4.2cm}+\s{_{ri}\over^{4\tau_2}}\s d\tau\wedge\s
+\s d\tau\wedge\da_\tau\s+\s{_{i}\over^{2\pi\kappa}}\s d\tau
\wedge\Delta_u\bigg)\NOm\bigg)\wedge\s\s\overline{\NOm}\s
\label{alla}
\qqq
where to obtain the last equality we have used the
relations (\ref{rella}).
The latter expression may be simplified if we notice
that
\qq
&{_i\over^{4\pi\kappa}}\s d\tau\wedge\sum_j\da_{u^j}
\bigg(\ee^{\s{_{\pi \kappa}
\over^{2\tau_2}}|u-\bar u|^2}\bigg((\da_{u^j}+{_{\pi\kappa}
\over^{\tau_2}}(u^j-\bar u^j)\bigg)\NOm\wedge\s\overline{\NOm}
\bigg)\s=\s\ee^{\s{_{\pi \kappa}
\over^{2\tau_2}}|u-\bar u|^2}\hs{1.9cm}&\cr
&\times\bigg(\bigg(
{_{\pi i\kappa}\over^{4\tau_2^2}}|u-\bar u|^2
d\tau\wedge+{_{i}\over^{2\tau_2}}
d\tau\wedge\sum_j(u^j-\bar u^j)\da_{u^j}
+{_i\over^{4\pi i\kappa}}\s d\tau\wedge
\Delta_u+{_{ri}\over^{4\tau_2}}
d\tau\wedge\bigg)\NOm\bigg)\wedge\overline{\NOm}\hs{0.5cm}&
\label{equal}
\qqq
with $\s\s\ee^{\s{_{\pi \kappa}
\over^{2\tau_2}}|u-\bar u|^2}\bigg((\da_{u^j}+{_{\pi\kappa}
\over^{\tau_2}}(u^j-\bar u^j)\bigg)\NOm\wedge\s\overline{\NOm}
\s\s$ being a $(2r)$-form well defined on $\CN$.
It follows that the fiber-wise integral of both sides of
eq.\s\s(\ref{equal}) vanishes and, consequently, that
\qq\non
\da_\tau\s\parallel \Psi\parallel^2&=&
{\rm const.}\s\s\tau_2^{-r/2}
\int\limits_{\CT_\tau^{\s r}}
\ee^{\s{_{\pi \kappa}
\over^{2\tau_2}}|u-\bar u|^2}\bigg(
\bigg(\da_\tau\s+\s{_i\over^{4\pi\kappa}}\s\Delta_u\bigg)
\NOm\bigg)\wedge\s\overline{\NOm}\cr
&&\hs{-1.3cm}=\s{\rm const.}\s\s\tau_2^{-r/2}
\int\limits_{\CT_\tau^{\s r}}
\ee^{\s{_{\pi \kappa}
\over^{2\tau_2}}|u-\bar u|^2}\s\s\overline{\theta(u)}\s\left(
\da_\tau\s+\s{_i\over^{4\pi\kappa}}\s\Delta_u\right)\theta(u)
\s\s d^{2r}u\s=\s\left(\Psi\s,\s\nabla_\tau\s\Psi\right)
\qqq
which proves the unitarity of the KZB connection for
the case with no insertions in a way that maybe was not the simplest
(two straightforward integrations by parts would do)
but which has the virtue of generalizing to the case
with arbitrary insertions (modulo convergence problems).
\vskip 0.3cm

To treat the general case, we shall apply the formula
(\ref{uint}) to the differential
form $\s\eta\s=\s\ee^{\s{_{\pi \kappa}
\over^{2\tau_2}}|w-\bar w|^2}\left\vert
\ee^{-{1\over\kappa}S}\s\NOm\right\vert^2\s$
with $\NOm$ given by eq.\s\s(\ref{dress})
and satisfying the generalization of relations (\ref{rella})
with $du^j$ replaced by $dw^j$.
As before, we obtain, denoting by $\da'$ the holomorphic
exterior derivative in all but $u$ directions,
\qq\non
&&\da\parallel \Psi\parallel^2\s=\s
{\rm const.}\s\s\tau_2^{-r/2}
\int\limits_{UY_{\tau,\un{z}}}
(\da+{_{ri}\over^{4\tau_2}}\s d\tau\wedge)\left(
\ee^{\s{_{\pi \kappa}
\over^{2\tau_2}}|w-\bar w|^2}\left\vert\s
\s\NOm\s\right\vert^2\right)\cr
&&=\s{\rm const.}\s\s\tau_2^{-r/2}
\int\limits_{UY_{\tau,\un{z}}}
\ee^{\s{_{\pi \kappa}
\over^{2\tau_2}}|w-\bar w|^2}\bigg(\bigg(
{_{\pi i\kappa}\over^{4\tau_2^2}}|w-\bar w|^2\s
d\tau\wedge\s+\s{_{\pi\kappa}\over^{\tau_2}}
\sum_j(w^j-\bar w^j)\s dw^j\wedge
\s+\s{_{ri}\over^{4\tau_2}}\s d\tau\wedge\cr
&&\hs{8cm}+\s \da'\s+\s\sum_j du^j\wedge\da_{u^j}\bigg)\NOm\bigg)
\wedge\s\overline{\NOm}\s\hs{0.5cm}\cr
&&=\s{\rm const.}\s\s\tau_2^{-r/2}
\hs{-0.2cm}\int\limits_{UY_{\tau,\un{z}}}
\hs{-0.15cm}\ee^{\s{_{\pi \kappa}
\over^{2\tau_2}}|w-\bar w|^2}\bigg(\bigg(
{_{\pi i\kappa}\over^{4\tau_2^2}}|w-\bar w|^2\s
d\tau\wedge\s+\s{_{i}\over^{2\tau_2}}
\s d\tau\wedge\sum_j(w^j-\bar w^j)\s\da_{u^j}\s
+\s{_{ri}\over^{4\tau_2}}\s d\tau\wedge\cr
&&\hs{0.5cm}+\s \da'\s+\s{_{i}\over^{2\pi\kappa}}\s d\tau
\wedge\Delta_u\s-\s{_1\over^{\kappa}}\sum_n\la^j_n\s
dz_n\wedge
\da_{u_j}\s+\s{_1\over^{\kappa}}\sum_s\alpha^j_s\s dy_s
\wedge\da_{u_j}\bigg)\NOm\bigg)
\wedge\s\s\overline{\NOm}\s.
\qqq
A large part of the right hand side vanishes by
integration by parts with use of the relation
generalizing (\ref{equal}):
\qq
&{_i\over^{4\pi\kappa}}\s d\tau\wedge\sum_j\da_{u^j}
\bigg(\ee^{\s{_{\pi \kappa}
\over^{2\tau_2}}|w-\bar w|^2}\bigg((\da_{u^j}+{_{\pi\kappa}
\over^{\tau_2}}(w^j-\bar w^j)\bigg)\NOm\wedge\s\overline{\NOm}
\bigg)\s=\s\ee^{\s{_{\pi \kappa}
\over^{2\tau_2}}|w-\bar w|^2}\hs{1.75cm}&\cr
&\times\bigg(\bigg(
{_{\pi i\kappa}\over^{4\tau_2^2}}|w-\bar w|^2
d\tau\wedge+{_{i}\over^{2\tau_2}}
d\tau\wedge\sum_j(w^j-\bar w^j)\da_{u^j}
+{_i\over^{4\pi i\kappa}}\s d\tau\wedge
\Delta_u+{_{ri}\over^{4\tau_2}}
d\tau\wedge\bigg)\NOm\bigg)\wedge\overline{\NOm}\hs{0.5cm}&
\label{equag}
\qqq
and we obtain
\qq
\da\parallel \Psi\parallel^2&=&
{\rm const.}\s\s\tau_2^{-r/2}
\int\limits_{UY_{\tau,\un{z}}}
\ee^{\s{_{\pi \kappa}
\over^{2\tau_2}}|w-\bar w|^2}\bigg(\bigg(
\da'\s+\s{_{i}\over^{4\pi\kappa}}\s d\tau
\wedge\Delta_u\s-\s{_1\over^{\kappa}}\sum_n\la^j_n\s dz_n\wedge
\da_{u_j}\cr
&&\hspace{7cm}+\s{_1\over^{\kappa}}\sum_s\alpha^j_s\s dy_s
\wedge\da_{u_j}\bigg)\NOm\bigg)
\wedge\s\s\overline{\NOm}\cr
&&\hs{-1.4cm}=\s{\rm const.}\s\s\tau_2^{-r/2}
\int\limits_{UY_{\tau,\un{z}}}
\ee^{\s{_{\pi \kappa}
\over^{2\tau_2}}|w-\bar w|^2}\bigg(\bigg(
\da'\s+\s{_{i}\over^{4\pi\kappa}}\s d\tau
\wedge\Delta_u\s-\s{_1\over^{\kappa}}\sum_n\la^j_n\s dz_n\wedge
\da_{u_j}\cr
&&\hspace{1.6cm}+\s{_1\over^{\kappa}}\sum_s\alpha^j_s\s dy_s
\wedge\da_{u_j}\bigg)\s\ee^{-{1\over\kappa}S}\s
\lle\s\Omega\s,\s\theta\s\rle\bigg)
\wedge\s\s\overline{\ee^{-{1\over\kappa}S}\s
\lle\s\Omega\s,\s\theta\s\rle}\ d^{2r}u\s.
\label{duj}
\qqq
The $\da_{u^j}$-terms may be transformed with use of relation
\qq
\sum_n\sum\limits_{i=1}^{K_n}\beta^j_{n,i}\s d(y_{n,i-1}-y_{n,i})\wedge
\Omega\s=\s{_i\over^{2\pi}}\s d\tau\wedge\da_{u^j}\Omega\s,
\label{arel}
\qqq
in the notations of eq.\s\s(\ref{Omeg}) and with $y_{n,0}\equiv z_n\s,$
\s which follows easily form the definitions of the forms
$\Omega$ and $\omega_q$. Eq.\s\s(\ref{arel}) may be rewritten as
\qq
\bigg(\sum_n\la^j_n\s dz_n\s-\s
\sum_s\alpha_s^j\s dy_s\bigg)\wedge\Omega\s=\s
{_i\over^{2\pi}}\s d\tau\wedge\da_{u^j}\Omega\s-
\s\sum_ndz_n\wedge h^j_{_{(n)}}\Omega\s,
\label{brel}
\qqq
with the contragradient action of $h^j_{_{(n)}}$ on the
$\un{V}^*$-valued form $\Omega$. The last relation, upon
substitution to eq.\s\s(\ref{duj}), yields
\qq\non
\da\parallel \Psi\parallel^2&=&
{\rm const.}\s\s\tau_2^{-r/2}
\int\limits_{UY_{\tau,\un{z}}}
\ee^{\s{_{\pi \kappa}
\over^{2\tau_2}}|w-\bar w|^2}\s
\left\vert\ee^{-{1\over\kappa}S}\right\vert^2
\bigg(\da'\lle\s\Omega\s,\s\theta\s\rle
\s-\s{_1\over^\kappa}\s\da S\wedge\lle\s\Omega\s,\s\theta\s\rle\cr
&&\hs{1.5cm}+\s{_i\over^{4\pi\kappa}}\s d\tau\wedge\Delta_u
\lle\s\Omega\s,\s\theta\rle\s-\s{_i\over^{2\pi\kappa}}
\s d\tau\wedge\sum_j\da_{u^j}\left(\lle\s\da_{u^j}\Omega\s,
\s\theta\rle\right)\cr
&&\hs{2.5cm}+\s{_1\over^{\kappa}}\sum_n
dz_n\wedge\da_{u^j}\s(\s\lle\s h^j_{_{(n)}}\Omega\s,
\hs{0.02cm}\s\theta
\s\rle\s)\bigg)\wedge\s\overline{\Omega}\ d^{2r}u\cr
&=&{\rm const.}\s\s\tau_2^{-r/2}
\int\limits_{UY_{\tau,\un{z}}}
\ee^{\s{_{\pi \kappa}
\over^{2\tau_2}}|w-\bar w|^2}\s
\left\vert\ee^{-{1\over\kappa}S}\right\vert^2
\bigg(\bigg\lle\da'\Omega\s-\s{_1\over^\kappa}\s\da S\wedge
\Omega\cr
&&\hs{3.3cm}-\s{_i\over^{4\pi\kappa}}\s d\tau\wedge\Delta_u
\Omega\s+\s{_1\over^{\kappa}}\sum_n
dz_n\wedge h^j_{_{(n)}}\da_{u^j}\Omega\s\hs{0.03cm},\
\theta\bigg\rle\cr
&&\hs{-0.4cm}+\s(-1)^K\bigg\lle\Omega\s\s,
\ \da'\theta\s+\s{_i\over^{4\pi\kappa}}\s d\tau\wedge\Delta_u\theta
\s-\s{_1\over^\kappa}\s\sum_ndz_n\wedge h^j_{_{(n)}}
\da_{u_j}\theta\bigg\rle
\bigg)\wedge\s\overline{\Omega}\ d^{2r}u\s.\hs{0.3cm}
\qqq
The crucial result are the following equalities:
\qq
\da'\Omega\s=\s0\quad\quad{\rm and}\quad\quad
(\da S)\wedge\Omega\s+\s d\tau\wedge H_0\hs{0.03cm}\Omega
\s+\s\sum_n dz_n\wedge H_n\hs{0.03cm}\Omega\s=\s0
\label{btr1}
\qqq
with the contragradient action of $H_n$'s.
The first of these equalities is a straightforward consequence
of the closedness of the forms $\omega_q$ from which $\Omega$
is built. The second, more technical one, has been announced
in \cite{FV} (as Prop.\s\s9). Using these relations, we finally
obtain
\qq\non
\da\parallel \Psi\parallel^2\s=\s
{\rm const.}\s\s\tau_2^{-r/2}
\int\limits_{UY_{\tau,\un{z}}}
\ee^{\s{_{\pi \kappa}
\over^{2\tau_2}}|w-\bar w|^2}\s
\left\vert\ee^{-{1\over\kappa}S}\right\vert^2\s
(-1)^K\s\s\lle\s\Omega\s\s,
\ \da'\theta\s+\s{_1\over^\kappa}\s d\tau\wedge H_0
\s\theta\s\cr+\s{_1\over^\kappa}\s\sum_ndz_n\wedge H_n\s\theta\s\rle
\bigg)\wedge\s\overline{\Omega}\ d^{2r}u\s=\s
(\s\Psi\s,\s\s\nabla\Psi\s)
\qqq
which proves the unitarity of the KZ connection w.r.t.
the scalar product (\ref{scpr}) modulo the control of
convergence of the integrals.
\vskip 0.9cm

\nsection{Bethe Ansatz}
\vskip 0.5cm

The basic algebraic relations (\ref{btr}) responsible
for the unitarity of the KZ connection turn out
to be a disguised (and compact) form of the Bethe Ansatz
solution of the eigenvalue problem for the Gaudin
Hamiltonians (\ref{Gaud}). This was remarked in \cite{Bab}
in the context of contour integral representations
for the solutions of the KZ equations and developed
further in \cite{BabFl,RV} and in \cite{FFR} where
a relation between the Bethe Ansatz and the Wakimoto realization
of the highest weight modules of Kac-Moody algebras was explained.
\vskip 0.2cm

The elementary fact is that if, for fixed $\un z$, a configuration
$\un y$ of $K$ non-coincident points in the plane satisfies
the equations
\qq
\da_{y_s}S^0(\un{z},\un{y})\s=\s0\s,\quad s=1,\dots,K\s,
\label{BA0}
\qqq
then the second relation of (\ref{btr}) reduces to
\qq\non
\sum_n dz_n\wedge\left(\da_{z_n}S^0(\un{z},\un{y})\s
+\s H^0_n(\un{z})\right)
\Omega^0(\un{z},\un{y})\s=\s0\s.
\qqq
The last equation gives the Bethe Ansatz solution for the common
eigenvectors of the operators $H^0_n(\un{z})\s$
acting by the contragradient representation in $\un{V}^*$:
\qq
\left(\da_{z_n}S^0(\un{z},\un{y})\s+\s H^0_n(\un{z})\right)
\s G^0(\un{z},\un{y})\ =\ 0
\label{BA01}
\qqq
in the notations of eqs.\s\s(\ref{G0},\ref{F0}).
The conditions (\ref{BA0}) required for eq.\s\s(\ref{BA01})
to hold have the explicit form
\qq
\sum_n\lle\la_n,\alpha_s\rle\s{1\over{z_n-y_s}}\s=\s\sum_{s'\not=s}
\lle\alpha_{s'},\s\alpha_s\rle\s{1\over{y_{s'}-y_s}}\s.
\label{BA02}
\qqq
\vskip 0.3cm

The above genus zero story has its elliptic counterpart,
as remarked in \cite{Etin-1} for the case of $G=SU(2)$,
see also \cite{FG}, and in \cite{FV} for general
simple groups. For fixed $\tau,\un{z}$ and for
$\un{y}$ satisfying the equations
\qq
\da_{y_s}S(\tau,\un{z},\un{y})\s=\s0\s,\quad s=1,\dots,K\s,
\label{BA1}
\qqq
or explicitly, with $\rho\equiv\vartheta_1'/\vartheta_1$,
\qq
\sum_n\lle\la_n,\alpha_s\rle\s\rho(z_n-y_s)\s=\s\sum_{s'\not=s}
\lle\alpha_{s'},\s\alpha_s\rle\s\rho(y_{s'}-y_s)\s,
\label{BA12}
\qqq
the second relation of (\ref{btr1}) reduces to
\qq
\bigg(d\tau\wedge\left(\da_{\tau}S(\tau,\un{z}.\un{y})\s+\s
H_0(\tau,\un{z})\right)\s+\s\sum_n dz_n
\wedge\left(\da_{z_n}S(\tau,\un{z},\un{y})\s
+\s H_n(\tau,\un{z})\right)\bigg)\s\cr
\times\ \Omega(\tau,u,\un{z},\un{y})\ =\ 0\s.
\label{njn}
\qqq
Operators $H_n(\tau,\un{z}),\ n\geq 0,$
act in the space of meromorphic functions $f$ of variable
$u\in\Nh^\NC$ taking values in $\un{V}^*_0$, obeying the
periodicity conditions
\qq
f(u+q^\vee)\s=\s f(u)\s,
\label{pr}
\qqq
with poles possible on the hyperplanes $\lle u,\alpha\rle\in\NZ
+\tau\NZ$\s. Eq.\s\s(\ref{njn}) gives the Bethe Ansatz solutions
for the common eigenvectors of $H_n$'s in that space
\qq\non
&&\left(\da_{\tau}S(\tau,\un{z},\un{y})\s\s \ +\s H_0(\tau,\un{z})\right)
\s G(\tau,u,\un{z},\un{y})\ =\ 0\s,\cr
&&\left(\da_{z_n}S(\tau,\un{z},\un{y})\s+\s H_n(\tau,\un{z})\right)
\s G(\tau,u,\un{z},\un{y})\ =\ 0
\qqq
in the notations of eqs.\s\s(\ref{F},\ref{G}).
\vskip 0.3cm

A slightly modified version of the above argument allows to
extend this construction and also to
find Bethe Ansatz eigenvectors of $H_n$'s acting in the space
of meromorphic $\un{V}^*_0$-valued functions satisfying
the twisted boundary conditions
\qq
f(u+q^\vee)\s=\s\ee^{\s\lle\xi,q^\vee\rle}\s f(u)
\label{prt}
\qqq
with $\xi\in\Nh^\NC$. Denoting
\qq\non
\tilde\Omega\s\equiv\s\ee^{\s\lle u,\xi\rle}\s\Omega\s,\quad\ \
\tilde S\s\equiv\s S\s+\s{_1\over^{4\pi i}}\s|\xi|^2\tau
\s-\s\sum_n\lle\xi,\la_n\rle z_n\s+\s\sum_s\lle\xi,\alpha_s\rle
y_s
\qqq
and combining the second equality
of (\ref{btr1}) with the relation (\ref{brel}), we obtain
\qq
(\da\tilde S)\wedge\tilde\Omega\s+\s d\tau\wedge H_0\hs{0.03cm}
\tilde\Omega\s+\s\sum_n dz_n\wedge H_n\hs{0.03cm}\tilde\Omega\s=\s0
\label{btr2}
\qqq
which results in the eigenvalue equations for the twisted-periodic
Bethe Ansatz eigenvectors:
\qq\non
&&\left(\da_{\tau}\tilde S(\tau,\un{z},\un{y})\s
\s\ +\s H_0(\tau,\un{z})\right)
\s\ee^{\s\lle u,\xi\rle}\s G(\tau,u,\un{z},\un{y})\ =\ 0\s,\cr
&&\left(\da_{z_n}\tilde S(\tau,\un{z},\un{y})\s+\s H_n(\tau,\un{z})\right)
\s\ee^{\s\lle u,\xi\rle}\s G(\tau,u,\un{z},\un{y})\ =\ 0
\qqq
holding provided that $\da_{y_s}\tilde S=0\s$, \s i.e.\s\s that
\qq
\sum_n\lle\la_n,\alpha_s\rle\s\rho(z_n-y_s)
\s+\s\lle\xi,\alpha_s\rle\s=\s\sum_{s'\not=s}
\lle\alpha_{s'},\s\alpha_s\rle\s\rho(y_{s'}-y_s)
\s.
\label{tBA12}
\qqq
In particular, for $\xi$ in the weight lattice $P$,
we obtain more periodic Bethe Ansatz eigenvectors.
\vskip 0.3cm

As explained in refs.\s\s\cite{Etin} or \cite{FV},
in the special case
of $G=SU(n)$ and one insertion of the $nm$-fold symmetric power
of the fundamental representation, operator $H_0$ acting
on the $\un{V}^*_0$-valued functions reduces to the elliptic
Calogero-Sutherland-Moser multi-body operator and
the above techniques were used in \cite{FV,FV1} to obtain its
Bethe Ansatz diagonalization. The above case
is a quantization of the very same elliptic Hitchin system
which appeared in the effective low energy description
of supersymmetric Yang-Mills theories \cite{DonWitt}.
For the $SU(2)$ case, one recovers this way \cite{Etin-1}
the classical Hermite's results about
the Lam\'{e} operator \cite{WW}.

\end{document}